\documentclass[useAMS,usenatbib,twocolumn]{mn2e}
\pdfoutput=1
\usepackage{amssymb}
\usepackage{amsmath}
\usepackage{natbib}
\usepackage[pdftex]{graphicx}
\usepackage{subfigure}
\usepackage{booktabs}

\begin{document}

\title[Magnetic buoyancy and flux pumping]
{Magnetic buoyancy instabilities in the presence of magnetic flux pumping at the base of
  the solar convection zone}
      \author[A.J. Barker, L.J. Silvers, M.R.E. Proctor \& N.O. Weiss]{Adrian J. Barker$^{1,3,}$\thanks{E-mail:
	  adrianjohnbarker@gmail.com}, Lara J. Silvers$^2$, Michael R.E. Proctor$^1$ \& Nigel O. Weiss$^1$, \\
	$^1$Department of Applied Mathematics and Theoretical
	Physics, University of Cambridge, Centre for Mathematical
        Sciences, \\ Wilberforce Road, Cambridge CB3 0WA, UK \\
        $^2$Centre for Mathematical Science, City University London,
        Northampton Square, London EC1V 0HB, UK \\
         $^3$Center for Interdisciplinary Exploration and Research in Astrophysics (CIERA) \& Dept. of Physics and Astronomy, \\ Northwestern University, 2145 Sheridan Rd, Evanston, IL 60208, USA. }

\date{Accepted 2012 April 24. Received 2012 April 23; in original form 2011 August 25}

\pagerange{\pageref{firstpage}--\pageref{lastpage}} \pubyear{2012}

\maketitle

\label{firstpage}

\begin{abstract}
We perform idealised numerical simulations
of magnetic buoyancy instabilities in three dimensions, solving the equations of
compressible magnetohydrodynamics in a model of the solar
tachocline. In particular, we study the effects of including a highly simplified model of
magnetic flux pumping in an upper layer (``the convection zone") on magnetic
buoyancy instabilities in a lower layer (``the upper parts of the radiative
interior -- including the tachocline"), to study these competing flux
transport mechanisms at the base of the convection zone. The results of the
inclusion of this effect in numerical simulations of the
buoyancy instability of both a preconceived magnetic slab and of a
shear-generated magnetic layer are presented. In the former, we find that if we
are in the regime that the downward pumping velocity is comparable with the Alfv\'{e}n
speed of the magnetic layer, magnetic flux pumping is able to hold
back the bulk of the magnetic field, with only small pockets of strong field able
to rise into the upper layer. 

In simulations in which the magnetic
layer is generated by shear, we find that the shear velocity is not
necessarily required to exceed that of the pumping (therefore the kinetic
energy of the shear is not required to exceed that of the overlying convection), 
for strong localised pockets of magnetic field to be produced which can rise
into the upper layer. This is because magnetic flux pumping acts
to store the field below the interface, allowing it to be amplified
both by the shear, and by vortical fluid motions, until pockets of field can achieve sufficient strength to rise into
the upper layer. In addition, we find that the interface
between the two layers is a natural location for the production of
strong vertical gradients in the magnetic field. If these gradients
are sufficiently strong to allow the development of magnetic
buoyancy instabilities, strong shear is not necessarily required to
drive them (c.f.\ previous work by Vasil \& Brummell).
We find that the addition of magnetic flux pumping appears to be able to
assist shear-driven magnetic buoyancy in producing strong flux concentrations that
can rise up into the convection zone from the radiative interior. 
\end{abstract}


\section{Introduction}
The Sun is observed to possess a large-scale predominantly toroidal
field at the surface, which exhibits cyclic magnetic activity, as manifested
by the sunspot cycle. Since the period of these cycles (approximately $11$ yr) is
much shorter than the ohmic diffusion time for a primordial solar
magnetic field ($10$ Gyr), it seems inescapable that we require the action of a
dynamo which can generate magnetic fields to explain this behaviour. Since 
dynamos are generally assisted by differential rotation, it is believed that
the tachocline, a region of strong radial (and somewhat weaker latitudinal) shear at the
base of the convection zone of the Sun, plays an important role in
this process, by generating toroidal field from poloidal
field\footnote{This is the so-called $\Omega$ effect of mean-field
  electrodynamics \citep{Moffatt1978}.} \citep{TobiasWeiss2007}. The bulk of the solar toroidal field
is likely to be stored just below
the convection zone, both because the action of convective turbulence has
been found to rapidly pump magnetic field into the nonturbulent
regions beneath (e.g.\ \citealt{SpiegelWeissNature1980};
\citealt{Tobias2001}), and also because the short rise time of buoyant magnetic flux
tubes would pose problems for the storage of
toroidal field at any depth in the convection zone \citep{Parker1975}.

Sunspots and other intense flux elements comprising active regions on
the solar surface are thought to be produced by the instability of this
large-scale toroidal field stored beneath the convection
zone. A prime candidate for the instability mechanism is  magnetic
buoyancy. This can be explained in brief as the 
idea that a horizontal magnetic field $\mathbf{B}$ can support heavy gas
above by virtue of the pressure ($|\mathbf{B}|^{2}/2\mu_{o}$) that it
exerts. If it is in thermal equilibrium, the gas in the region of the field will
therefore rise, since it is lighter
than its surroundings. A horizontal magnetic field that decreases with height can
render the fluid top-heavy, which is liable to
instability. 

Previous work has studied the instability of a field that
decreases smoothly with height, in which case the unstable modes can
be either two-dimensional ``interchange modes'' or three-dimensional
``undular modes'', of which the latter are usually dominant (see, for
example, the following review articles: \citealt{HughesProctorReview1988};
\citealt{Hughes2007}). If the field
is discontinuous and consists of a slab of
horizontal magnetic field sandwiched by nonmagnetic gas in a
convectively stable atmosphere, then the instability is of Rayleigh-Taylor
type, and occurs for any strength of magnetic field (in the absence of
diffusion). In this case the
instability is a two-dimensional interchange mode, which involves no
bending of the field lines \citep{CH1988}. However, the nonlinear
evolution of the instability can generate three-dimensional arched
structures that qualitatively resemble those observed at the
surface. These arise through the interactions between vortices, 
which are primarily generated by Kelvin-Helmholtz instabilities (vorticity is also produced by baroclinicity, a linear effect associated with the initial instability) driven by the rising
magnetic ``mushrooms'' (\citealt*{Matthews1995b}; \citealt{Wissink2000}).
An important question concerns the scales of the rising magnetic structures. In the absence of diffusion the instability occurs at very small scales, and it is hard to see how such modes can lead to the large coherent structures seen at the surface. Larger length scales can be obtained either by using enhanced turbulent diffusion (and numerical computations are inevitably diffusive and show the same effect), or as a result of helicity in the magnetic field structures, for example through rotation.

In reality, the predominant source of the solar toroidal field is
likely to be the strong radial (and somewhat weaker latitudinal) shear
in the tachocline. This effect results from the variation in
the angular velocity in the solar interior, which can stretch any
poloidal field to produce toroidal field. While any poloidal field is
unlikely to be coherent in the tachocline\footnote{This is because a coherent poloidal
field which straddles the base of the convection zone would cause the
differential rotation of the convection zone to be imprinted onto the
radiative interior, which is not what we observe
(e.g.\ \citealt{MacGregor1999}; though also see
\citealt{Rogers2011}).}, it is certainly likely to
be present to some degree. Recent work has therefore begun to address
the problem of the generation of a toroidal magnetic layer through
shear, together with the resulting 
magnetic buoyancy instabilities of this layer (e.g.\ \citealt{BCC2002};
\citealt{CBCdynamo2003}; \citealt{VB2008}, hereafter VB08; \citealt{VB2009}, hereafter VB09;
\citealt*{SVBP2009}; \citealt*{SBP2009}, hereafter SBP09).

Of most relevance to this work, VB08 and SBP09 used numerical simulations in Cartesian geometry to
study the generation and subsequent instability of a horizontal
magnetic layer from an initially uniform vertical field. They found
that strong velocity shear, which is hydrodynamically unstable to
Kelvin-Helmholtz type shear instabilities, is
required for magnetic gradients to be sufficient for magnetic buoyancy instabilities to develop.
Since the shear in the tachocline is believed to be much weaker, and hydrodynamically
stable, this result appeared to provide a problem for the efficacy of this mechanism in
producing buoyant flux structures. 
However, it is known that radiative diffusion (with diffusivity $\kappa$)
below the convection zone is much more efficient at transporting heat than
ohmic diffusion (with diffusivity $\eta$) is at transporting magnetic flux. In the regime that
$\kappa \gg \eta$, it has long been known that double-diffusive
effects could allow magnetic buoyancy instabilities
(\citealt{Gilman1970}; \citealt{Acheson1979}; \citealt{SR1983})
by mitigating the stabilising stratification through the action of radiative
diffusion. It is therefore possible that a hydrodynamically stable
tachocline shear can produce a magnetic layer that is unstable to a
double-diffusive magnetic buoyancy instability. This was first
confirmed by the simulations of \cite*{SVBP2009}, who used
a similar setup to VB08, except in a parameter regime which
favoured double-diffusive instabilities (see also \citealt{Lara2011}). They showed that
magnetic buoyancy instabilities of a shear-generated magnetic layer
are possible in the double-diffusive regime, which is relevant
for the Sun. In this paper we use the simplest model in which to investigate magnetic buoyancy instabilities, and do not study double diffusive effects, while recognising that a more complicated model should be considered in due course.

Whatever the mechanism by which these structures are produced, they then rise into the solar convection zone. The earlier calculations on shear-induced buoyancy did not in general include the action of the turbulent convection in this region. It is likely that the convection zone plays a crucial role in the solar dynamo process by helping to return flux of an appropriate orientation to the shear region so that the dynamo cycle can be completed. The effect of anisotropic and inhomogeneous turbulence on flux transport has been known for some time
 (\citealt{DrobYuf1974}; \citealt{Arteretal1982};
\citealt{Arter1983}; \citealt{GallProc83}; \citealt{Tao1998}). The principal effect is to transport horizontal magnetic flux down the gradient of
turbulent intensity (\citealt{Zeldovich1957}; \citealt{Moffatt1983}). 
In the presence of a stable
layer beneath, which is absent in convective turbulence, magnetic flux can be
transported and stored in this layer (\citealt{Tobias1998}; \citealt{Tobias2001};
\citealt{DorchNordlund2001}; \citealt{Ossendrijver2002}; \citealt{KitRud2008}). Numerical
simulations of turbulent penetrative compressible convection
show that magnetic flux is rapidly transported into the underlying stable layer on a
convective (and not a diffusive) timescale by strong downflowing plumes (\citealt{Tobias1998};
\citealt{Tobias2001}; \citealt{DorchNordlund2001}), and that  only the strongest field is able to counteract this effect and rise into the upper unstable layer. This phenomenon is robust
and its physical basis is straightforward: in a stratified compressible convecting fluid there is a sharp distinction between
rapidly falling and gently rising plumes \citep{Weiss2004}.

The purpose of this paper is to carry out a pilot study of the effects of the turbulent pumping on the evolution of the buoyancy instability. The computations  referred to above show clearly that such an interaction is meaningful, with flux being transported not only on the smallest scales of the convection but on much larger scales too, so that there is a `mean' effect.  Ideally we would wish to study the interaction of magnetic buoyancy instabilities
with the overlying convective flows in a direct numerical simulation,
such as an extension of those presented in SBP09. However this is computationally challenging at the present time. Furthermore, we wish to understand the basic physics of the interaction of the atmosphere and the growing instabilities, without getting distracted by the complexities of the actual convection. Thus we choose to begin by constructing a model problem that captures the effect of the turbulence on the largest scales of the buoyancy modes by means of a mean-field ansatz.
Clearly this simplification does not allow consideration of the largest scales of motion which are comparable or larger than the important buoyancy modes.
Nor does it properly treat the small scales of the
buoyancy which may be comparable with or smaller
than those of the convection. Nonetheless the simple
ansatz used does in our view capture important elements of the interaction, and gives a useful guide
to more detailed calculations in the future. Note
that our approach will not be strictly accurate at the base
of the convection zone, where convection exists on
a wide range of scales (e.g.~\citealt{Miesch2008}).
Averaging over the largest convection cells would
then be meaningful only for the largest scales of the
buoyant magnetic field. However, these models are
not designed to accurately simulate the conditions
at the base of the convection zone of the Sun. Instead, they are designed to provide a simple phenomenological picture of the magnetic pumping of
large-scale fields resulting from small-scale convection, which can hopefully provide a complementary
perspective to simulating the convection directly.


In keeping with the philosophy of modelling processes at the simplest non trivial level, we represent the magnetic pumping via mean-field
electrodynamics as the antisymmetric part of the $\alpha$-tensor
(e.g.\ \citealt{Radler1968}; \citealt{Moffatt1983}), i.e.,
$\alpha_{ij} = \alpha \delta_{ij} - \epsilon_{ijk}\gamma_{k}$. This
contributes an additional advective velocity to the induction equation
for the mean magnetic field through the term $\nabla \times
\left(\boldsymbol{\gamma} \times \mathbf{B}\right)$, where
$\boldsymbol{\gamma}$ is a turbulent pumping velocity, which arises
from a gradient in the turbulent intensity of the flow.  

In this paper, we therefore study the effects of this mechanism on magnetic buoyancy instabilities by implementing a
$\gamma$-pumping effect in numerical simulations of
standard magnetic buoyancy.

We first extend previous calculations of the buoyancy
instability of a preconceived magnetic slab (\citealt{CH1988};
\citealt*{Matthews1995b}) by including the additional effect of
magnetic pumping in an upper layer (``the convection zone''). This is designed to 
crudely mimic the addition of one of the most important effects of the
convection zone on the nonlinear evolution of magnetic buoyancy
instabilities of this magnetic slab. Though this is a
very simple model, it is worthwhile to extend previous simulations
through the addition of only this effect, to study in detail its
influence on the problem. Later on, we extend previous calculations in which this
magnetic layer is generated through shear acting on a vertical field,
by including magnetic pumping in an upper layer. This should allow us
to isolate the most important effect of an overlying convection zone
on the nonlinear outcome of these instabilities.

The structure of the paper is as follows. We first describe the
numerical setup of our simulations starting with a preconceived
magnetic slab in \S \ref{model1}. In this section we also discuss the
simple model of magnetic pumping that we adopt. We then present the results of these
simulations in \S \ref{results1}. The numerical setup for the problem
with shear is described in \S \ref{modelwithshear}, and the
corresponding results in \S \ref{resultswithshear}. Finally, in \S
\ref{discussion} we present a discussion of the results and a summary
of those which we deem to be important. 

\section{Numerical Model (without shear)}
\label{model1}

We adopt a Cartesian box with coordinates $(x,y,z)$, which represents
a plane section of the tachocline region of the Sun. The entire domain extends from
$z=0$ (top) to $z=d$ (bottom) in the vertical, and extends from
$x,y=0$ to $x,y=\lambda_{x,y} d$ in the horizontal, where
$\lambda_{x,y}$ will be specified later. We identify $xz$ as the
poloidal plane (where $-\mathbf{e}_{z}$ is considered to be the radial direction), and
$y$ as the toroidal (azimuthal) coordinate.
An infinite slab of uniform toroidal
magnetic field $\mathbf{B} = B_{0}\mathbf{e}_{y}$ is placed in
the region $z\in [z_{1},z_{2}]$, which is sandwiched by nonmagnetic
gas. The initial state is static, with a
linear temperature distribution $T=T_{0}\left(1+\theta z/d \right)$, and is piecewise
polytropic (with index $m$), being computed under
the assumption that the total (gas plus magnetic) pressure is continuous
at each magnetic interface. This is an unstable equilibrium
configuration. If a small perturbation is imposed on the system, the density discontinuity at
$z=z_{1}$ is unstable to a Rayleigh-Taylor type instability, which we
excite by adding small random thermal perturbations to the top of the magnetic layer.

We assume the fluid is a perfect gas with constant shear viscosity
$\mu$, thermal conductivity $\kappa$, magnetic diffusivity $\eta$ and
specific heats $c_{p}$ and $c_{v}$ (which define the gas constant $\mathcal{R} = c_{p}-c_{v}$). The standard equations of mass,
momentum, entropy and magnetic induction, together with the equation
of state, can be non-dimensionalised by scaling lengths with the
depth of the layer $d$, temperatures with the initial temperature at the
upper surface $T_{0}$, densities with the initial density at the upper surface
$\rho_{0}$, and magnetic fields with the magnitude of the initial
magnetic field $B_{0}$ (e.g.\ \citealt*{Matthews1995a}). We also use $d/\sqrt{\mathcal{R}T_{0}}$ as the
unit of time, which corresponds to the sound travel time, using the
isothermal sound speed at the top of the layer. Using this
non-dimensionalisation, equations governing the temporal evolution of the density $\rho$,
velocity $\mathbf{u}$, temperature $T$ and magnetic field $\mathbf{B}$ read
\begin{eqnarray}
\label{1}
\partial_{t}\rho + \nabla \cdot \left(\rho \mathbf{u}\right) = 0
\end{eqnarray}
\begin{eqnarray}
\label{2}
\nonumber
\partial_{t}\left(\rho\mathbf{u}\right)\hspace{-0.2cm} &=&\hspace{-0.2cm} 
-\nabla \left(p+\frac{F}{2}|\mathbf{B}|^{2}\right) + \nabla \cdot
\left(F \mathbf{B} \mathbf{B} - \rho \mathbf{u}\mathbf{u} + \sigma
  C_{K} \boldsymbol{\mathrm{\tau}} \right) \\ && +\theta (m+1)\rho \mathbf{e}_{z}
\end{eqnarray}
\begin{eqnarray}
\label{3}
\nonumber
\partial_{t}T &=& -\mathbf{u}\cdot \nabla T - (\gamma -1)T \nabla \cdot
\mathbf{u} + \frac{\gamma C_{K}}{\rho} \nabla^{2}T  \\ &&+
\frac{C_{K}(\gamma-1)}{\rho}\left(\frac{\sigma}{2} ||\boldsymbol{\mathrm{\tau}}||^{2} +
  F\zeta_{0} |\mathbf{J}|^{2}\right)
\end{eqnarray}
\begin{eqnarray}
\label{4}
\partial_{t}\mathbf{B} =
\nabla \times \left[\mathbf{u}\times \mathbf{B} - \zeta_{0}C_{K}
  \mathbf{J}\right] + \mathbf{G}
\end{eqnarray}
\begin{eqnarray}
\label{5}
\nabla \cdot \mathbf{B} = 0,
\end{eqnarray}
where the current $\mathbf{J} = \nabla \times \mathbf{B}$, the
equation of state is $p = \rho T$, and the viscous stress tensor is
\begin{eqnarray} 
\tau_{ij} = \partial_{j}u_{i} + \partial_{i}u_{j} -
\frac{2}{3}\delta_{ij}\partial_{k}u_{k},
\end{eqnarray}
and $\mathbf{G}$ will be specified later.

These equations contain seven dimensionless parameters, which, together with the initial and boundary
conditions, completely determine the evolution of the system. These are the
polytropic index $m$, the temperature gradient
$\theta$, and the ratio of specific heats $\gamma=c_{p}/c_{v}$,
together with the Prandtl number
\begin{eqnarray}
\sigma = \frac{\mu c_{p}}{\kappa},
\end{eqnarray}
the ratio of magnetic to thermal diffusivity at the top of the layer
\begin{eqnarray}
\zeta = \frac{\eta \rho_{0}c_{p}}{\kappa},
\end{eqnarray}
the dimensionless thermal diffusivity
\begin{eqnarray}
C_{K} = \frac{\kappa}{\rho_{0}c_{p}d\sqrt{\mathcal{R}T_{0}}},
\end{eqnarray}
and the dimensionless field strengh
\begin{eqnarray}
F = \frac{B_{0}^{2}}{\mathcal{R}T_{0}\rho_{0}\mu_{0}}.
\end{eqnarray}
The last quantity is related to the plasma $\beta$, which is the ratio of gas to magnetic
pressure, by $F = 2/\beta$. Note that the number of pressure scale heights in the domain is given
by $N_{p} = (m+1)\ln (1 + \theta)$. We always take
$\gamma=\frac{5}{3}$, as is appropriate for a monatomic ideal gas, and
will from now on reuse the symbol $\gamma$ to represent magnetic flux pumping, which
we will define in \S~\ref{gammapumping}.

Eqs.\ \ref{1}--\ref{5} are solved subject to the boundary conditions that 
\begin{eqnarray}
&& u_{z} = \partial_{z}u_{x} = \partial_{z}u_{y}=0 \;\;\;\;\;\;\;\;\;\;\;\; \mathrm{at}
\;\;\; z=0,1, \\
&& T=1  \;\;\; \mathrm{at} \;\;\; z=0, \;\;\; \partial_{z}T=\theta \;\;\; \mathrm{at} \;\;\; z=1.
\end{eqnarray}
These conditions represent impermeable, stress-free boundaries at the
top and bottom of the computational domain. All horizontal (magnetic and nonmagnetic) boundary
conditions are periodic. The mass flux and
mechanical energy flux thus vanish on the boundaries, and the imposed
heat flux is the only flux of (non-magnetic) energy into and out of
the system. The vertical
magnetic boundary conditions are
\begin{eqnarray}
B_{x}=B_{y}=0 \;\;\; \mathrm{at} \;\;\; z=0,1.
\end{eqnarray}
The field is therefore ensured to be vertical at the top and bottom
boundaries. Note that any imposed horizontal field
can leave the domain, since a gradient of these fields can
exist at the boundaries. This choice of boundary conditions is somewhat artificial. However, the dynamics 
in the region not close to the vertical boundaries should
only be weakly influenced by them.

The numerical method used to solve the above system of equations
is a parallel hybrid finite-difference/pseudospectral code, where spatial
derivatives are calculated in Fourier space for the horizontal
directions and fourth-order finite-differences in the vertical
direction (upwind derivatives being used for the advection terms). 
Time integration is performed by an explicit third-order Adams-Bashforth method.
The equations solved, and the numerical method
used are discussed in more detail in \cite*{Matthews1995a} and
\cite{Bushby2005}, for example. 
The code has been thoroughly tested, particularly on problems of magnetoconvection and
magnetic buoyancy. 

We simulate a box that is elongated in the direction of the initial
field, by choosing $\lambda_{x}=1$ and $\lambda_{y}=4$, so that the
vortex tube instability observed by \cite*{Matthews1995b} is allowed
to develop and produce three-dimensional structure.
We use a spatial resolution of $128\times 128\times
200$, except where specified otherwise.

\subsection{Downward magnetic pumping}
\label{gammapumping}
We add the following term into
Eq.~\ref{4} to represent the effects of turbulent pumping of
the magnetic flux from the upper ``convective'' regions, of the form
\begin{eqnarray}
\mathbf{G} = \nabla \times \left(\boldsymbol{\gamma} \times \mathbf{B} \right),
\end{eqnarray}
with 
\begin{eqnarray}
\label{gammaprofile}
\boldsymbol{\gamma}  = \frac{\gamma_{m}}{2}\left[1+\tanh \left[(\Delta z_{i})^{-1}\left(z_{i}-z\right)\right]\right]\mathbf{e}_{z}.
\end{eqnarray}
The vertical profile is designed to represent a region with a
uniform nonzero value (``the convection zone'') that
smoothly goes to zero (``the radiation zone'')
below a depth $z_{i}$ (``the radiative-convective interface''). We plot an example
of such a profile in Fig.~\ref{gammaprofile}, using a set of typical
values for the various parameters. This is a purely downward pumping velocity, which should act to
prevent magnetic field with horizontal strengths smaller than
\begin{eqnarray}
|\mathbf{B}_{h}| = B_{eq} \equiv \gamma_{m}\sqrt{\frac{\rho(z_{i})}{F}},
\end{eqnarray}
from rising into the upper layer. The evolution of the system after
the onset of buoyancy instabilities will therefore
depend on the parameter (noting that we initially take $B_{y}=1$)
\begin{eqnarray}
  M_{\gamma} = \gamma_{m}\sqrt{\frac{\rho
      (z_{i})}{F}} = \frac{B_{eq}}{B_{y}},
\end{eqnarray}
which is an Alfv\'{e}nic Mach number for the $\gamma$-pumping.
If $M_{\gamma} \lesssim 1$, we would expect some of the rising
field to be able to overcome the downward pumping and rise into the
upper layer. Alternatively, if $M_{\gamma} \gtrsim 1$, we would expect
the field to be too weak to overcome the downward pumping, and
$\gamma$ should therefore act as a lid which will hold down the field,
unless the field can be locally amplified sufficiently that $|\mathbf{B}_{h}| >
B_{eq}$.

\begin{figure}
  \begin{center}
    \subfigure{\includegraphics[width=0.52\textwidth]{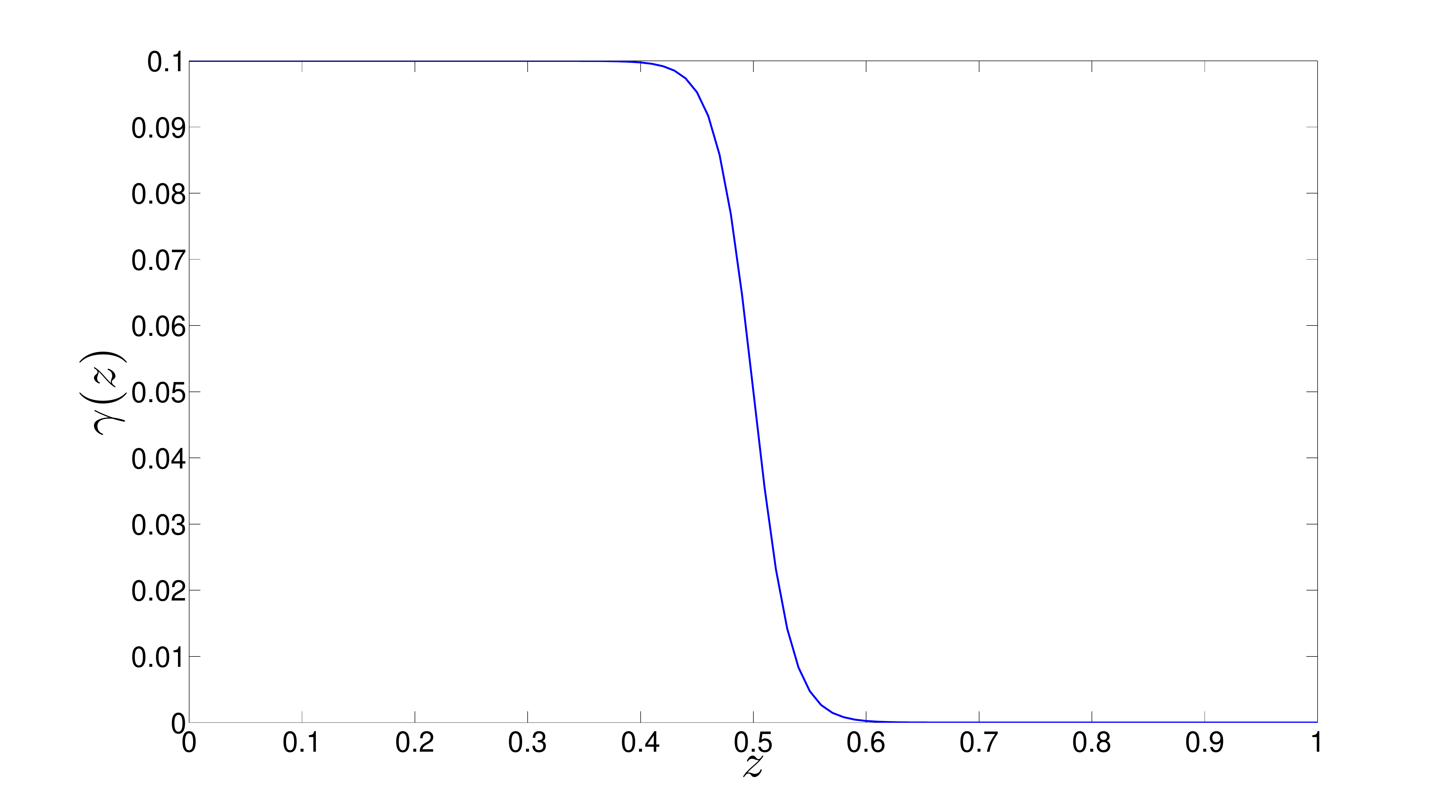} } 
  \end{center}
  \caption{Typical profile of magnetic pumping with $z_{i}=0.5, (\Delta
    z_{i})^{-1}=30$ and $\gamma_{m}=0.1$.} 
  \label{gammaprofile}
\end{figure}

The interpretation that we will adopt is that
the Cartesian box represents a large
horizontal section at the base of the convection zone. In this case,
the spatial scale of the convection cells is considered to be much
smaller than the horizontal size of the box. We
can then reasonably assume a separation of scales and define an
average over the small-scale convection
cells. Our vertical scale is such that the upper layer
($z< z_{i}$) represents a significant fraction
of the lower half of the convection zone, and $\gamma$
can be considered a ``mean-field''
pumping effect. In this interpretation, the simulated magnetic field, $\mathbf{B}$, is always considered to be a ``mean field", with the horizontal length scales of the magnetic field being much larger than that of the (unresolved) convection. It is then reasonable to consider $\gamma$ to act with equal strength on all of the resolved horizontal scales of $\mathbf{B}$ in our simulation, to a first approximation. For numerical reasons, it is essential to include nonzero diffusivities of momentum, heat and magnetic field. This means that the resolved diffusive lengthscales in the simulation are, by construction, larger than the horizontal scales of the unresolved small-scale convection. We consider the actual diffusive lengthscales to be much smaller than those of the unresolved convection. If desired, the simulated diffusivities can be considered to represent turbulent diffusivities resulting from the unresolved convection. However, we only include them for numerical reasons (though we will later include their effects in our discussion, for completeness), since we do not set out to study the effect of turbulent diffusivities on magnetic buoyancy.

Convective turbulence is also likely to pump the scalar fields of density and
temperature, in addition to the magnetic field. If the turbulence is
incompressible and only varies in one direction, such scalar pumping
vanishes (leaving only anisotropic diffusion) because the pumping velocity is divergence-free
\citep{Moffatt1983}, and when it exists it is, in general, not
simply related to the velocity of magnetic pumping \citep{CHP88}. For compressible
turbulence, it can be shown that an additional mean advection term arises (e.g.\ \citealt{VA1997}),
which may not necessarily vanish when the intensity of the turbulence
varies only with height, as in our case. However, for simplicity, and since this effect 
is not simply related to $\gamma$, we choose to neglect pumping of scalar
fields throughout this paper.

\subsection{Parameters adopted}
\label{noshearparams}
Convective flows in the lower parts of the convection zone are likely
to be highly subsonic, with mach numbers inferred to be in the range
$10^{-4}-10^{-2}$ (e.g.\ \citealt{Ossendrijver2003}; \citealt{Jones2010}). In the simulations of
\cite{Tobias2001}, the magnetic pumping effect resulting from
turbulent compressible convection occurs on a convective
timescale. We might therefore expect $\gamma$ to be subsonic,
with velocities at most comparable to the convective flows,
constraining $\gamma_{m} \ll 1$.

In the tachocline, $\beta \sim 10^{7}$ (e.g.\ \citealt{TH2004}),
which constrains $F \ll 1$. We must choose a much smaller $\beta$ for
these calculations than in reality to speed up the initial
instability. This is because the fastest growing Rayleigh-Taylor type mode has a growth
rate which scales with $\beta^{-1}$ \citep{CH1988}. We wish to study the
nonlinear evolution of the instability in the presence of
$\gamma$-pumping in the upper layer for a range of values
of $M_{\gamma}$ either side of unity. To do this we fix a value of $F=0.01$
and vary $\gamma_{m}$. This fixes the growth rate
and horizontal wavenumber of the initial instability, so that we can
better isolate the consequences of varying the strength of the downward pumping.
The parameter values adopted for these simulations are summarised in Table \ref{table1}. 

At the base of the convection zone
the diffusivities are ordered such that $1 \gg \kappa \gg \eta \gg
\nu$ (\citealt{Gough2007}; \citealt{Jones2010}). We respect this ordering by choosing $1 \gg
\zeta_{0} \gg \sigma$ and $C_{K} \ll 1$, though we do take much larger
diffusivities than are present in the Sun, as is required if we are to
run fully resolved simulations with a sensible run time. The stratification that we adopt has approximately three
pressure scale heights within the domain. This is designed to roughly
correspond with a region straddling the base of the convection zone
\citep{JCDMT2007}. Note, however, that the upper layer in which $\gamma$-pumping is
present also has subadiabatic stratification (since we fix $m>
1/(1-\gamma) = 3/2$), unlike in the convection zone of the Sun. We
have also performed simulations in which the initial stratification changes
from adiabatic to subadiabatic when $z>0.5$, which might be more
appropriate for the Sun. However, the results of these simulations did
not differ significantly from those with subadiabatic stratification
throughout the box; therefore we only
discuss simulations with a single polytropic index in this paper, for
clarity.

\begin{table}
\begin{center}
\begin{tabular}{l  c  r} 
\hline 
 Parameter & Description & Value \\ 
\hline
$m$ & Polytropic index & 1.6 \\
$\theta$ & Thermal stratification & 2.0 \\
$C_{K}$ & Thermal diffusivity & 0.01 \\ 
$\zeta_{0}$ & Magnetic diffusivity & 0.01 \\
$\sigma$ & Prandtl number & 0.005 \\
$F$ & Magnitude of magnetic energy & 0.01 \\
$\gamma_{m}$ & Magnetic pumping strength & Various \\
$z_{i}$ & Bottom of pumping layer & 0.5 \\
$(\Delta z_{i})^{-1}$ & Width of transition region & 30.0 \\
$z_{1},z_{2}$ & Top \& bottom of magnetic slab & 0.6, 0.8 \\
$\lambda_{x},\lambda_{y}$ & Box horizontal aspect ratio & $1,4$ \\
\hline
\end{tabular}
\end{center}
\caption{Parameter values for the discontinuous field cases.}
\label{table1}
\end{table}

\section{Results: instability of a magnetic layer in non-magnetic
  background with downward pumping}
\label{results1}

In this section we discuss the results of our simulations with a discontinuous magnetic layer initially in
magnetostatic equilibrium with its surroundings, including magnetic flux pumping in an upper
layer. The nonlinear breakup of such a layer in the absence of
flux pumping was studied by \cite{CH1988} in 2D, and
\cite*{Matthews1995b} and \cite{Wissink2000} in 3D. 

Since we begin with a discontinuous field, diffusion rapidly smears the
interface of the magnetic layer, though this does not significantly
influence the dynamics of the resulting buoyancy instabilities. 
The initial perturbation kinetic energy decays rapidly after
pushing the system from equilibrium, with the resulting buoyancy
instability being of Rayleigh-Taylor type, driven by the free
energy associated with the release
of gravitational potential energy stored in the initial state. This
instability develops by perturbing the upper interface of
the magnetic layer, eventually resulting in the formation of ``magnetic mushrooms''. The
most unstable mode has a horizontal wavenumber of approximately four when the initial
configuration is randomly
perturbed. The local rising of the field at the top of the layer
results in shear between field and field-free regions. This is subject
to Kelvin-Helmholtz instabilities that produce vortices 
at these interfaces, at the sides of the
mushrooms \citep{CH1988}. These vortices interact, and neighbouring vortex tubes
with opposite vorticity (from adjacent mushrooms) become
longtitudinally unstable to an analogue of the
(antisymmetric) Crow instability between neighbouring counter-rotating
vortex tubes \citep{Crow1970}. This produces three-dimensional structure in the
direction of the field, inducing arching of the rising
magnetic structures \citep*{Matthews1995b}.

\begin{figure*}
  \begin{center}
   \subfigure{\includegraphics[width=0.35\textwidth]{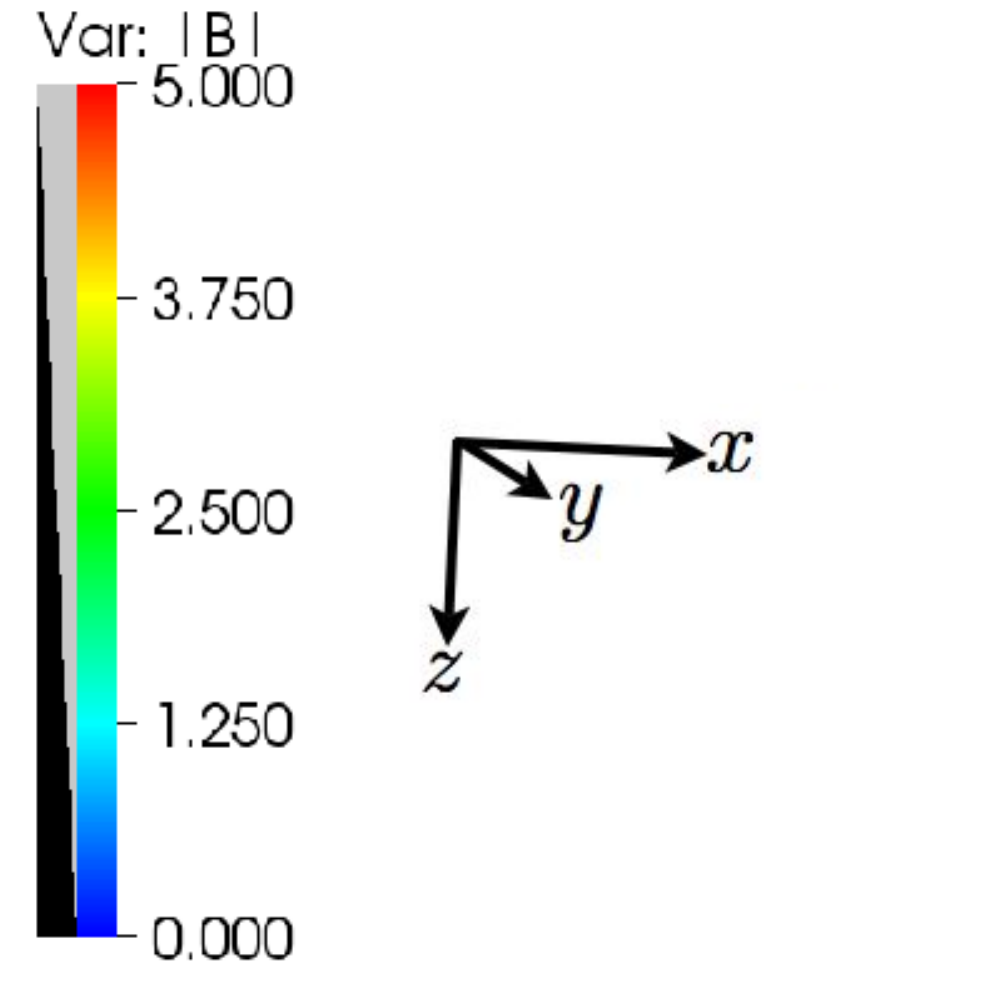} }
    \subfigure{\includegraphics[width=0.4\textwidth]{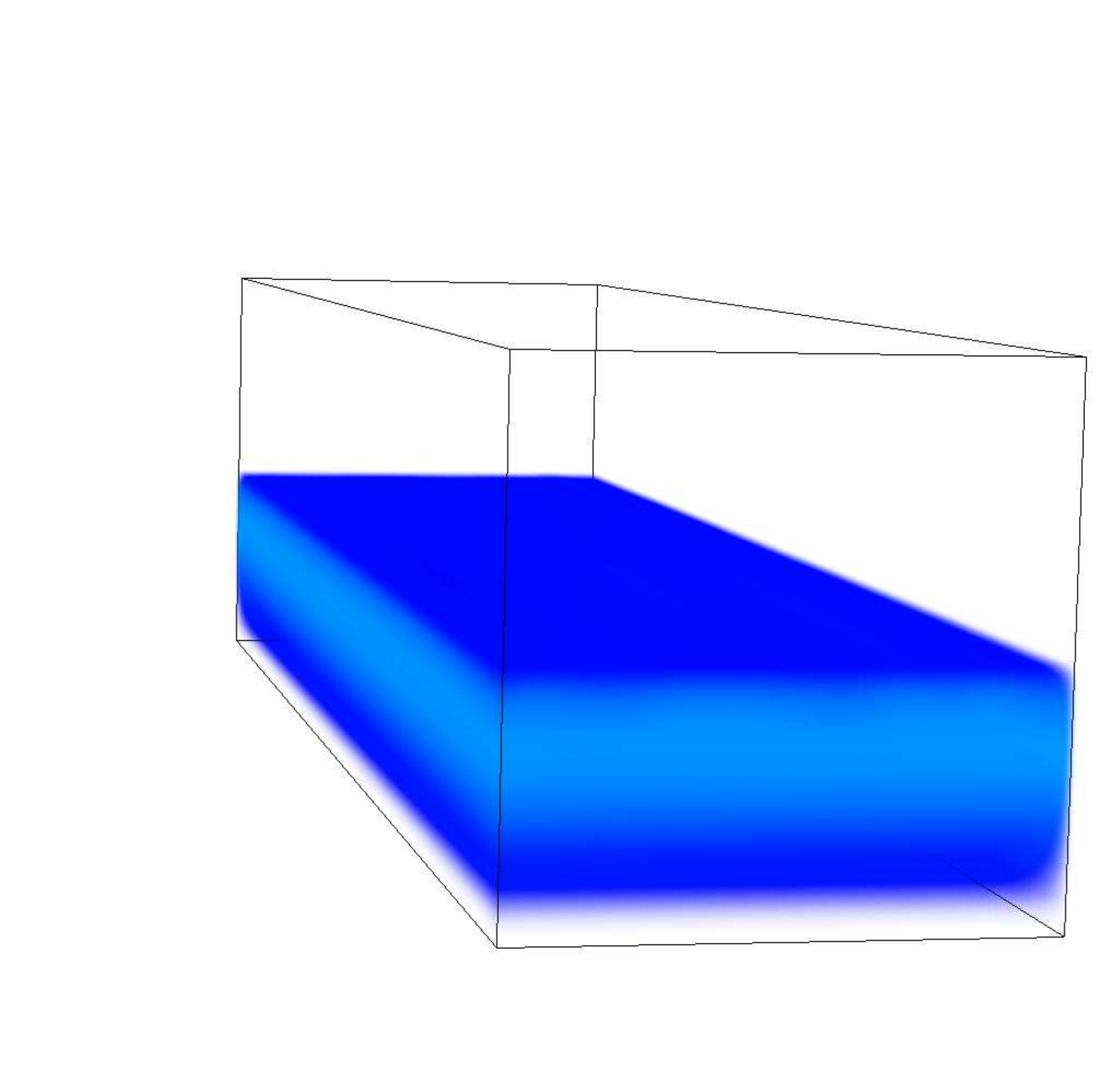} } \\
    \subfigure{\includegraphics[width=0.4\textwidth]{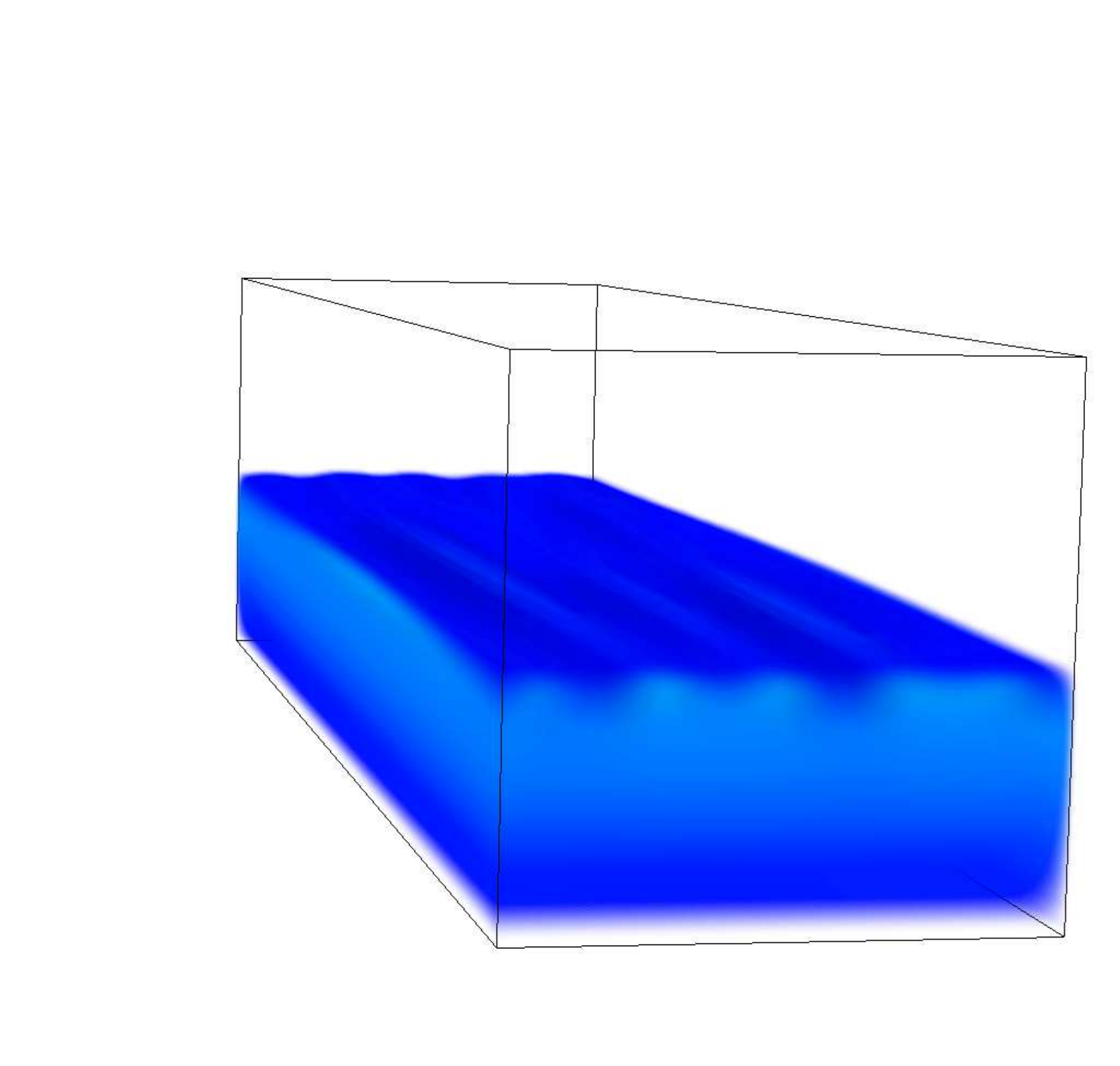} } 
 \subfigure{\includegraphics[width=0.4\textwidth]{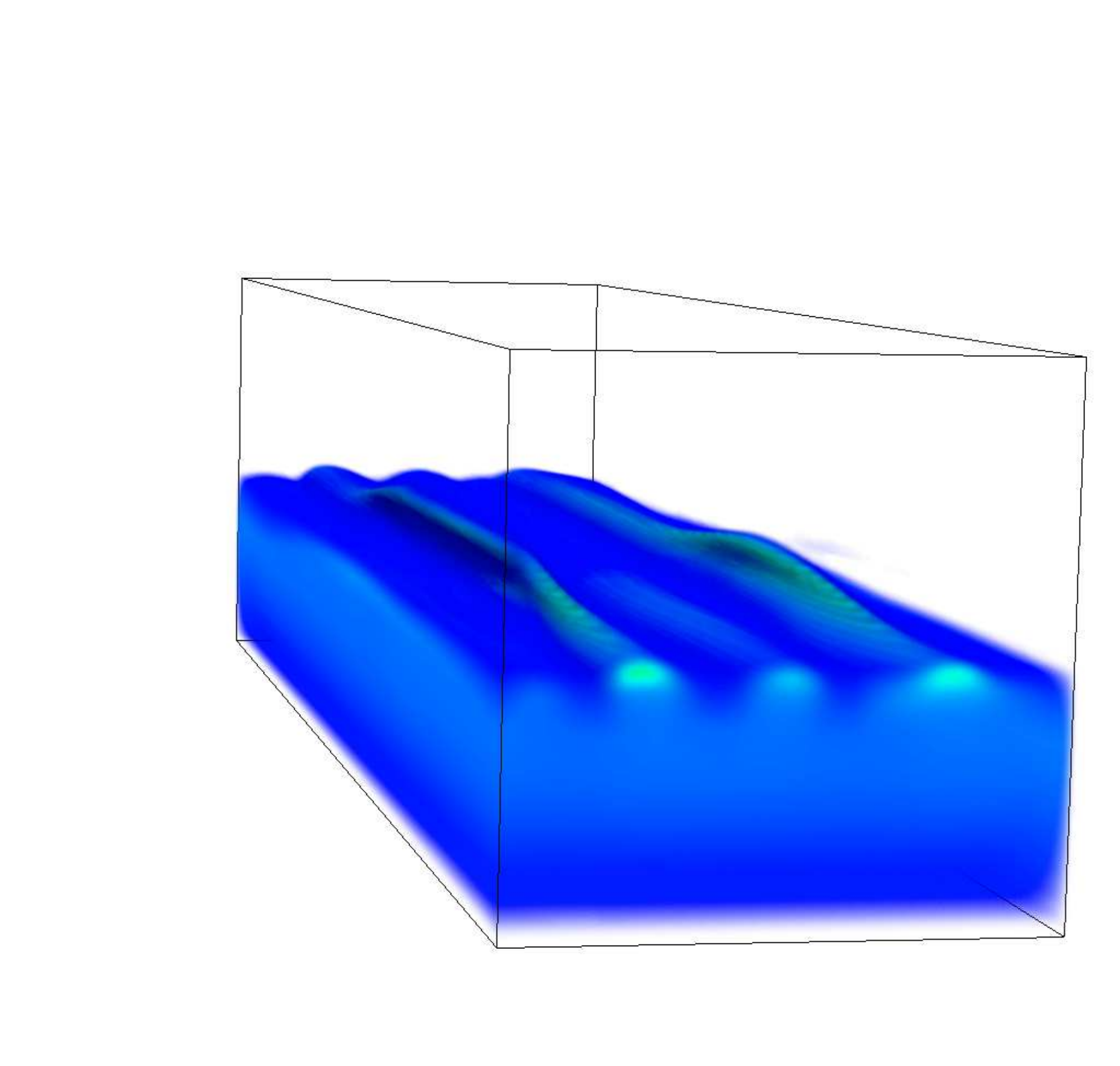} } \\
\subfigure{\includegraphics[width=0.4\textwidth]{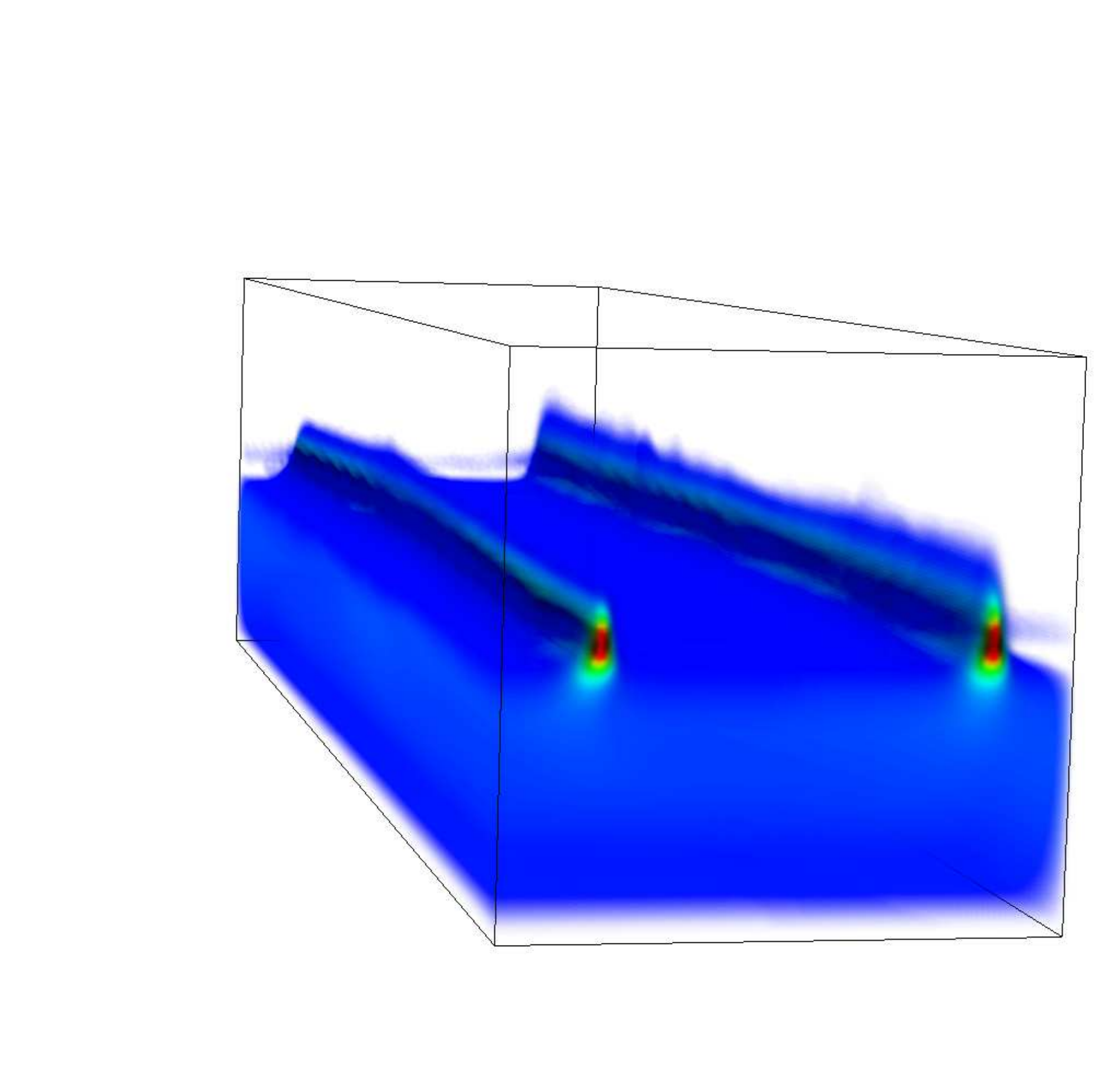} } 
\subfigure{\includegraphics[width=0.4\textwidth]{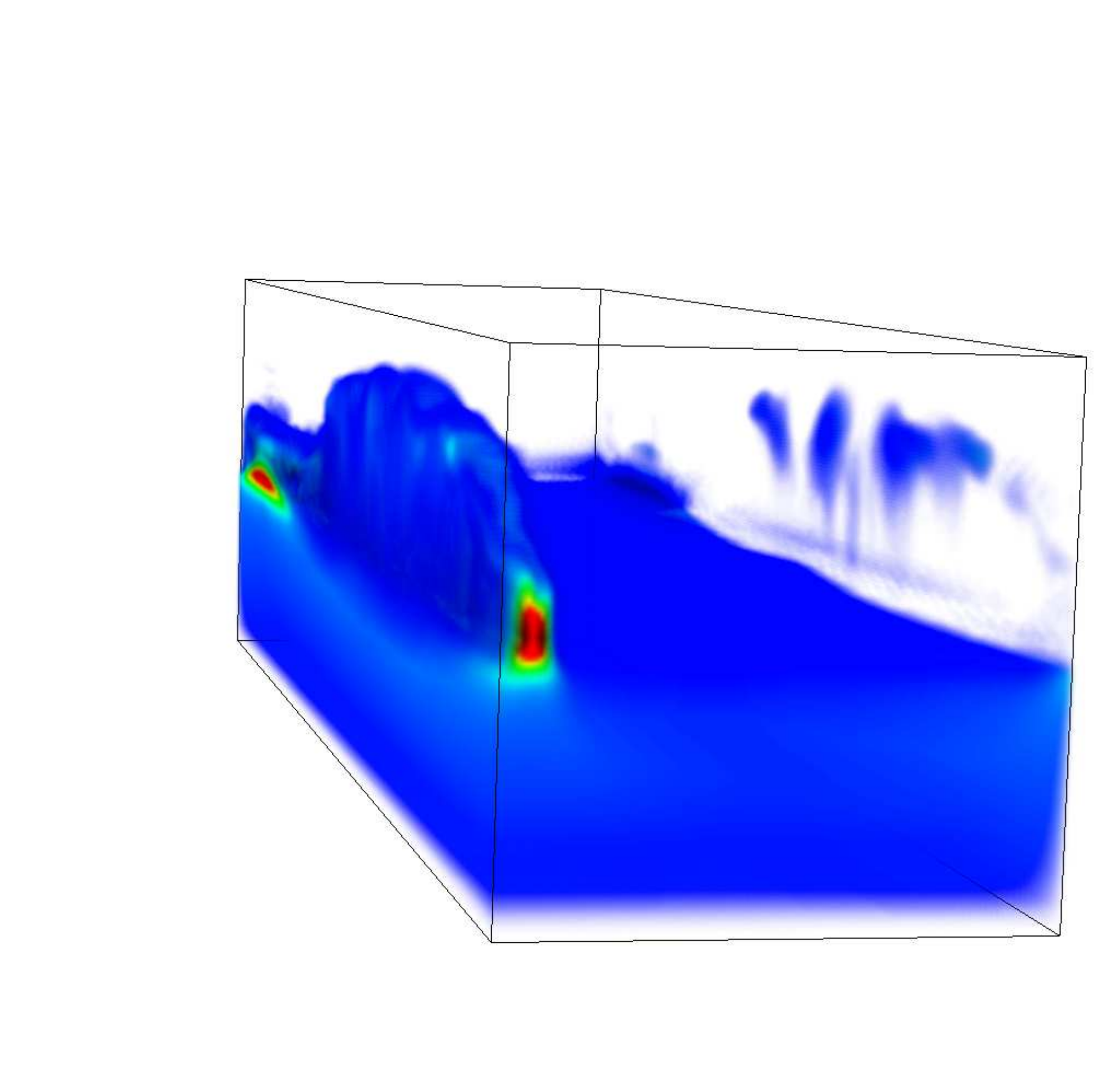} }
  \end{center}
  \caption{Volume renderings of $|\mathbf{B}|$ for a simulation with
    $M_{\gamma}=1$ at approximate times $t = 46, 98, 123, 267$ and $325$,
    respectively. This illustrates the
    temporal evolution of the magnetic field in a simulation of the
    buoyancy instability of a preconceived magnetic slab with magnetic
    pumping in the upper layer.
 } 
\label{DFvolume}
\end{figure*}

Volume rendering images illustrating the temporal evolution of our simulation with $M_{\gamma}=1$
for $|\mathbf{B}|$ are presented in Fig.~\ref{DFvolume}. So far what we have discussed is identical to the evolution described in
 \cite*{Matthews1995b} and \cite{Wissink2000}, which is what we would expect until 
 the field has risen far enough to reach the $\gamma$-interface. Once the
 buoyant magnetic structures reach $z\approx z_{i}$, the resulting evolution
 depends on the relative strength of the field compared to that which can be held down by the
 $\gamma$-pumping, i.e.\ whether $|\mathbf{B}_{h}|>B_{eq}$, which clearly depends on the value
 of $M_{\gamma}$ for the initial state. If $M_{\gamma} \ll 1$ downward pumping is unable to counteract the
upward transport of magnetic flux due to buoyancy. In this regime, the
rising magnetic structures are able to rise into the upper layer effectively
unhindered, and on horizontal scales comparable to the initial
instability. 

The more interesting regime is when $M_{\gamma} \gtrsim
1$, since downward pumping is then efficient at holding down the bulk of the
magnetic field, as must be the case in the Sun. However, localised
pockets of magnetic field with strengths
satisfying $|\mathbf{B}_{h}|>B_{eq}$ can still be produced. Several mechanisms
are responsible for this. 
One is a nonlinear effect, in which rising pockets of magnetic field generate
vortices, therefore the region
in the vicinity of $z_{i}$ is subject to complicated 
interactions between them. In some cases, these interactions are able
to concentrate magnetic flux below the
interface, primarily horizontally, to produce localised pockets of
strong field (we later illustrate this in a simulation with shear in
Fig.~\ref{SHKHMU2volume}). Two other effects which can locally concentrate
the field are the vertical variation of the $\gamma$-pumping, which can amplify field
through its nonzero divergence\footnote{This is a linear process and occurs because the effects of 
  $\gamma$-pumping in the induction equation are
\begin{eqnarray}
\left(\partial_{t} + \gamma\partial_{z}\right)B_{x,y} &=&
-B_{x,y}\partial_{z}\gamma = -\left(\nabla \cdot \gamma\right)
\mathbf{B}\cdot \mathbf{e}_{x,y} \\
\left(\partial_{t} + \gamma\partial_{z}\right)B_{z} &=& 0,
\end{eqnarray}
for which the horizontal components contain a forcing term as a result
of the nonzero divergence of $\gamma$. This leads to
some field amplification in vicinity of the interface between the two layers.
}, together with the effect of buoyancy instabilities below the
interface driving flux upwards into the interface. A combination of
these effects can produce 
localised peak fields that satisfy $|\mathbf{B}_{h}|>B_{eq}$, and
so are able to rise, even
if the layer as a whole is contained because $M_{\gamma} \gtrsim
1$.

Rising pockets of magnetic field are ultimately sheared apart by
interacting with the gas in the upper
layer and do not rise far as coherent structures. Since they are predominantly toroidal
structures, they are not transversally supported by magnetic tension,
and are therefore easily destroyed (see
e.g.\ \citealt{FalleHughes1998a}; \citealt{FalleHughes1998b}). 
Magnetic structures that are arched in the lower layer become straightened
as they rise up through the $\gamma$-transition region. This is
because, although $\gamma$ is horizontally
uniform, there exists a vertical transition region in which $\gamma$
increases with height. The peaks of the arches are therefore
pushed downwards more strongly than the troughs, which straightens the
field lines as they rise through this region. It must be noted that this is an artifact of our
adopted profile of magnetic pumping, and arched magnetic structures
could easily be produced if there were some horizontal variation in the
strength of $\gamma$. In reality, arched magnetic structures could
be produced either by the initial instability or its early nonlinear
evolution, or by the action of
turbulent motions in the convection zone on initially straight
structures, with the latter being observed by \cite{JouveBrun2009},
for example.

\begin{figure}
  \begin{center}
\subfigure{\includegraphics[width=0.5\textwidth]{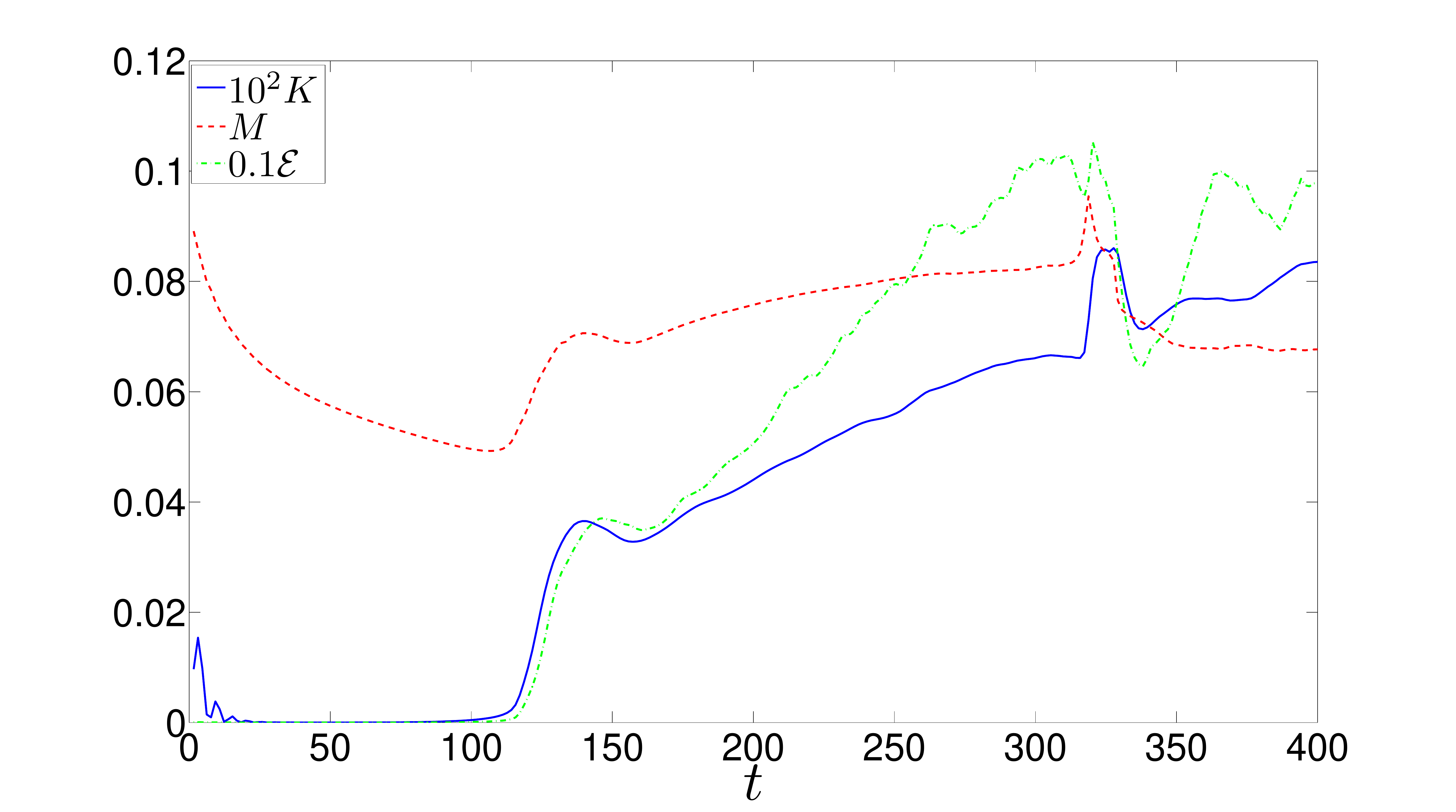} }  
  \end{center}
  \caption{Temporal evolution of the
    volume-integrated magnetic energy $M$, kinetic energy $K$ and
    enstrophy $\mathcal{E}$, in a simulation with $M_{\gamma}=1$,
    where these quantities have been scaled as listed in
    the legend.} 
\label{DFEnergy}
\end{figure}

During this simulation, the
potential energy of the initial configuration is transformed to
kinetic energy of the initial instability. Almost immediately, the
shearing motions of the field generate vortices through
Kelvin-Helmholtz instabilities.
Illustrated in Fig.~\ref{DFEnergy}, the integrated magnetic energy ($M = \int_{0}^{1} \langle
\frac{1}{2}|\mathbf{B}|^{2} \rangle dz$, where we denote a
horizontal average by $\langle X \rangle =
\int \int X dx dy$) initially builds until the
instability has developed and localised strong field pockets start to
rise into the upper layer (at
$t\approx 110$). Once this occurs, the subsequent generation of vorticity is illustrated by the increase
in kinetic energy ($K = \int_{0}^{1} \langle
\frac{1}{2}\rho|\mathbf{u}|^{2}\rangle dz$) 
and enstrophy ($\mathcal{E} = \int_{0}^{1} \langle \frac{1}{2}|\nabla \times
\mathbf{u}|^{2}\rangle dz$) within the computational domain
for $t\gtrsim 110$.
Afterwards, the initial instability dies out and
viscous and ohmic diffusion result in a slow decay of the integrated
energies. Since the influence of the (artificial) upper boundary becomes more important
as the simulation progresses, we only analyse the results until
$t\approx 400$.

\begin{figure}
  \begin{center}
\subfigure{\includegraphics[width=0.5\textwidth]{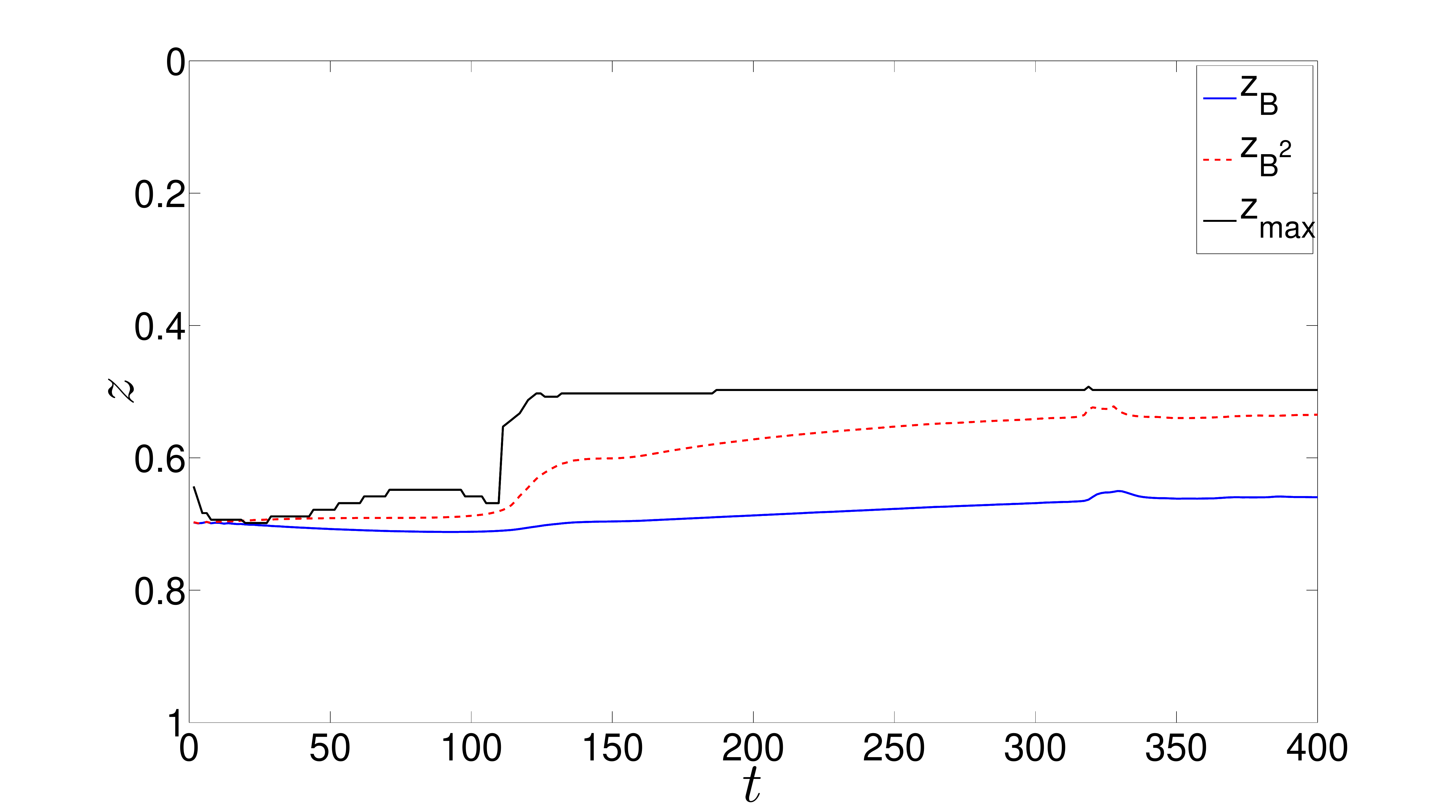} }
  \end{center}
  \caption{Temporal evolution of the peak magnetic field $z_{\mathrm{max}}$,
    the centre of magnetic field $z_{B}$, and the centre of
    magnetic energy $z_{B^{2}}$, in a simulation with $M_{\gamma}=1$.} 
\label{DFCentrefield}
\end{figure}

It is instructive to define several measures to describe the spatial
distribution of magnetic flux in the computational domain, and
therefore understand the relative efficiencies of magnetic buoyancy and magnetic
pumping. We can define the vertical location of the peak magnetic field, the centre of
magnetic field (e.g.\ \citealt{Wissink2000}; \citealt{Tobias2001}), and the centre of magnetic energy, respectively, as
\begin{eqnarray}
\label{measure1}
  z_{\mathrm{max}} &=& \underbrace{\mathrm{max}}_{\mathbf{|B|}} \left\{ z\right\}, \\
  z_{B} &=& \frac{\int_{0}^{1}z\langle B_{y}\rangle
    dz}{\int_{0}^{1}\langle B_{y} \rangle dz}, \\
  z_{B^{2}} &=& \frac{\int_{0}^{1}z\langle |\mathbf{B}|^{2} \rangle
    dz}{\int_{0}^{1}\langle |\mathbf{B}|^{2} \rangle dz}. 
\label{measure3}
\end{eqnarray}
We plot these measures as a function of time in Fig.~\ref{DFCentrefield},
where it can be seen that the peak field is located at a depth $z \approx z_{i}$. 
Note that the bulk of the field is held down in the
lower layer, since $z_{B} > z_{i}$. This is a result of both the
interaction between neighbouring vortices, which can act in concert to
hold down the bulk of the field, as has been previously observed by
\cite{CH1988}, together with $\gamma$-pumping in the upper layer.
Note that $z_{B^{2}}$ is located slightly higher than
$z_{B}$. This difference can occur when either $\mathbf{B}$ contains
appreciable $B_{x}$ or $B_{z}$ higher up, or alternatively if the field contains unsigned
magnetic field in this region. The latter can be produced by the induction of small-scale field
by vortices, which is most important at the top of the
transition region. The depth of the peak field moves around
chaotically because magnetic field is advected by vortical fluid motions
near the interface, though it always remains near $z_{i}$. 

\begin{figure}
  \begin{center}
    \subfigure{\includegraphics[width=0.5\textwidth]{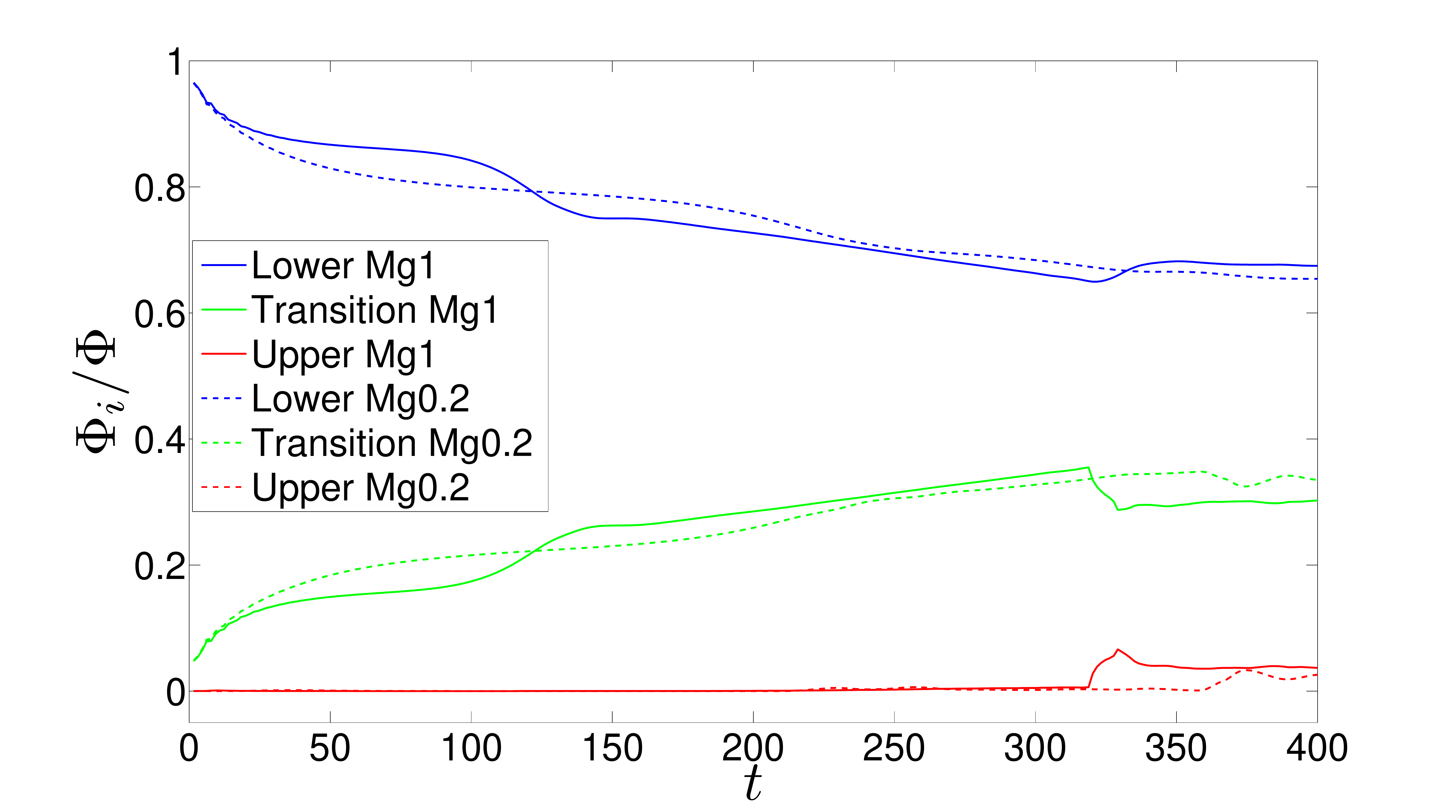} } 
  \end{center}
  \caption{Temporal evolution
    of the magnetic flux fraction contained within the lower layer $\Phi_{L}/\Phi$, the
    transition region $\Phi_{T}/\Phi$ and the upper layer
    $\Phi_{U}/\Phi$, in a simulation with $M_{\gamma}=1$ (solid lines)
    and $M_{\gamma}=0.2$ (dashed lines).} 
\label{DFflux}
\end{figure}

We define the total magnetic flux, as well as the magnetic flux contained
within the lower, transition and upper regions as
\begin{eqnarray}
\Phi &=& \int_{0}^{1} \langle B_{y} \rangle dz,  \\
\Phi_{U} &=& \int_{0}^{0.4} \langle B_{y} \rangle dz, \\
\Phi_{T} &=& \int_{0.4}^{0.6} \langle B_{y} \rangle dz \\
\Phi_{L} &=& \int_{0.6}^{1} \langle B_{y} \rangle dz.
\end{eqnarray}
We plot each of the last three normalised to the total $\Phi$ (which
is a time varying quantity, primarily due to ohmic dissipation) at each
time in Fig.~\ref{DFflux}. 
Magnetic pumping gradually pumps any flux out of the upper layer, competing
against localised breakouts (and ohmic diffusion). 
The fraction of the magnetic flux contained in the upper
layer is always less than approximately $5\%$ of the total initial flux, which reinforces the localised
nature of the breakouts. In addition, the field is primarily of a diffuse
nature in the upper layer because the coherence of the rising pockets
of flux are not maintained. Note that $\Phi_{U}/\Phi$ is slightly smaller in
the simulation with $M_{\gamma}=0.2$ than that with $M_{\gamma}=1$,
which might seem surprising. This is because the
amplification of field through the divergence of $\gamma$ near $z_{i}$
is weaker, whereas the bulk of the field is held down as efficiently by vortices.

To summarise the results of this section, we find that although the
nonlinear evolution of buoyancy instabilities of a preconceived
magnetic slab can produce
localised pockets of strong field which are able to rise up against
$\gamma$-pumping, the bulk of the field is held down in the lower
layer. This is a result of a combination of the interactions between
vortices and $\gamma$-pumping in the upper layer. A combination of
these effects might be responsible for holding down the bulk of the
solar toroidal field, allowing only localised breakouts into the
convection zone.

\section{Numerical Model with shear}
\label{modelwithshear}

The simulations described so far have assumed a preconceived
horizontal magnetic field
configuration in the initial state. From here on, we extend our study of magnetic pumping to
examine the effect of its addition on the generation of the horizontal magnetic layer by
the action of vertical shear on an 
initially uniform
vertical field, following VB08 and SBP09. This shear is designed to mimic the radial shear in the
tachocline. We do not consider any latitudinal (horizontal) shear because this
is inferred to be much weaker than the radial component (e.g.\ \citealt{JCDMT2007}). Our adopted
vertical shear profile is
\begin{eqnarray}
\mathbf{U}_{0}  = \frac{U_{m}}{2}\left[1+\tanh \left[(\Delta z_{s})^{-1}\left(z_{s}-z\right)\right]\right]\mathbf{e}_{y}.
\end{eqnarray}
We choose $U_{m} \ll 1$ so that this shear is subsonic, as is
appropriate for the tachocline. Unfortunately,
it is numerically feasible to simulate the tachocline shear only when
it is hydrodynamically unstable to a Kelvin-Helmholtz type
instability, in which case the instabilities are standard, i.e.\ not double-diffusive, instabilities. This is because capturing the double-diffusive instabilities would require very high resolution. This means that it is only possible to
perform a parameter survey when
$Ri = \mathrm{min}\left\{N^{2}/\left(\frac{d\mathbf{U_{0}}}{d
        z}\right)^{2}\right\} \lesssim \frac{1}{4}$, which is unlikely to
be valid in the Sun \citep{Gough2007}.
Nevertheless, the resulting
Kelvin-Helmholtz instabilities are suppressed when the
induced horizontal field strength becomes sufficiently large (e.g.\ SBP09). This
is usually well before the onset of any magnetic buoyancy instabilities, so it
should not significantly influence the dynamics that we are
interested in (except by producing inhomogeneities in the
magnetic layer along the direction of the induced field).

In previous work, VB08 and VB09 have studied a similar problem and
found some difficulties in maintaining the initial shear
profile. They used an applied stress, which is designed to counter viscous
decay, to maintain the shear. However, this is not able to maintain
the background shear in the presence of strong magnetic fields, and
the shear profile that remains at the end of their simulations did not
match that of the desired (initial) profile (see VB08 Fig.~13). This motivated us to
consider a different approach which will maintain the desired shear
profile and
might therefore be better at capturing the important effects of tachocline
shear on our problem (particularly for long duration simulations). Our approach is to decompose the velocity field into $\mathbf{u} =
\mathbf{u}^{\prime} + \mathbf{U}_{0}$, and consider $\mathbf{U}_{0}$ to be
steady, i.e.\ we consider the tachocline to be externally maintained. 
We neglect the back-reaction of $\mathbf{u}^{\prime}$ on
$\mathbf{U}_{0}$, and solve Eqs.\ \ref{1}--\ref{5} for
$\mathbf{u}^{\prime}$ instead of the total velocity field
$\mathbf{u}$. This has the advantage that
the underlying shear profile is
not affected by the magnetic buoyancy instabilities that we are interested
in studying\footnote{\cite{Kapyla2011} used another approach in which the
  horizontally-averaged velocity profile is relaxed back to the
desired shear flow over some prescribed timescale. We do not adopt
their approach since it is likely to
interfere unphysically with the desired dynamics of the buoyancy
instabilities.}.

We now outline the modifications to the numerical
model outlined in \S \ref{model1} which allow this problem to be
simulated. The initial state is now a (single) polytropic
layer permeated by a uniform vertical magnetic field $\mathbf{B} = B_{0}
\mathbf{e}_{z}$. In the absence of shear ($\mathbf{U}_{0}=\mathbf{0}$), this
is an equilibrium solution (neglecting diffusion). However, when our
adopted shear profile is present, the initial state is no longer an
equilibrium configuration because the vertical field is stretched by the shear to produce a
horizontal field. The instability of this field will
produce magnetic structures that buoyantly rise into the upper layer,
where they are acted upon by downward pumping, which is implemented as in \S \ref{gammapumping}.

We explicitly add additional terms into Eqs.\ \ref{1}--\ref{4} that
take into account the forcing of the flow by the background shear
$\mathbf{U}_{0}$. These terms are
\begin{eqnarray}
F_{\rho} &=& -\mathbf{U}_{0}\cdot \nabla \rho, \\
\mathbf{F}_{\mathbf{u}} &=& -\mathbf{U}_{0}\cdot \nabla \mathbf{u}^{\prime} - \mathbf{u}^{\prime}
\cdot \nabla \mathbf{U}_{0}, \\
F_{T} &=& -\mathbf{U}_{0}\cdot \nabla T, \\
\mathbf{F}_{\mathbf{B}} &=& \nabla \times \left( \mathbf{U}_{0}\times
  \mathbf{B} \right),
\end{eqnarray}
which must be added onto the right hand sides of Eqs.\ \ref{1}--\ref{4} (in that order),
where we have taken into account that $\nabla \cdot \mathbf{U}_{0} =
0$. 

\subsection{Parameters adopted}

The parameters adopted for our calculations with shear are outlined in
Table \ref{table2}, where it can be noted that most of the parameters remain
unchanged from the model in \S \ref{model1} that we have summarised in Table
\ref{table1}.

\begin{table}
\begin{center}
\begin{tabular}{l  c  r} 
\hline 
 Parameter & Description & Value \\ 
\hline
$m$ & Polytropic index & 1.6 \\
$\theta$ & Thermal stratification & 2.0 \\
$C_{K}$ & Thermal diffusivity & 0.01 \\
$\zeta_{0}$ & Magnetic diffusivity & 0.01 \\
$\sigma$ & Prandtl number & 0.005 \\
$F$ & Magnitude of magnetic energy & $10^{-5}$ \\
$\gamma_{m}$ & Magnetic pumping strength & Various \\
$z_{i}$ & Bottom of pumping layer & 0.5 \\
$(\Delta z_{i})^{-1}$ & Width of transition region & 30.0 \\
$U_{m}$ & Magnitude of shear & 0.1 \\
$z_{s}$ & Location of shear & 0.75 \\
$(\Delta z_{s})^{-1}$ & Width of shear region & 30.0 \\
$\lambda_{x},\lambda_{y}$ & Box horizontal aspect ratio & $1,4$ \\
\hline
\end{tabular}
\end{center}
\caption{Parameter values for the simulations with shear.}
\label{table2}
\end{table}

In the presence of shear, once magnetic field gradients become sufficient
for magnetic buoyancy instabilities to occur, the resulting evolution
will depend on the Alfv\'{e}nic Mach number of the shear-generated
magnetic layer. The strength of the magnetic field in the layer will
grow according to 
\begin{eqnarray}
B_{y} \approx \frac{B_{z}U_{m}}{2\Delta z_{s}}t,
\end{eqnarray}
until such a time as the layer has spread due to the slow propagation of Alfv\'{e}n
waves along the initial vertical field. The associated timescale for
this process is $t \approx \Delta z_{s}/(2v_{A})$, which is the
vertical Alfv\'{e}n travel time across half of the shear region.
In our units, the peak field is therefore $\mathrm{max}(B_{y}) \approx
U_{m}\sqrt{\frac{\rho (z_{s})}{F}}$. This means that our parameter $M_{\gamma}$
from \S~\ref{model1} is equivalent to
 \begin{eqnarray}
 M_{\gamma} = \frac{B_{eq}}{\mathrm{max}(B_{y})} = \frac{\gamma_{m}}{\mathrm{max}(B_{y})}\sqrt{\frac{\rho (z_{i})}{F}} \approx
 \frac{\gamma_{m}}{U_{m}}\sqrt{\frac{\rho (z_{i})}{\rho (z_{s})}}.
 \end{eqnarray}
The evolution of buoyantly unstable field which reaches the upper
layer will therefore depend on the parameter
\begin{eqnarray}
M_{U} = \frac{\gamma_{m}}{U_{m}},
\end{eqnarray}
which will play an analogous role to $M_{\gamma}$ from \S~\ref{model1}
(the density ratio is approximately unity). Noting that
$(1/2)\rho\gamma_{m}^{2}$ is likely to be comparable to (though
smaller than) the kinetic energy of the
convection (and remembering that $\gamma$ is not an
actual fluid velocity in our mean-field interpretation), $M_{U}^{2}$
is a measure of the ratio of the kinetic energy of
the convection to that of the shear. If $M_{U} \lesssim 1$, we
would expect shear to generate sufficiently strong horizontal magnetic fields for
the resulting rising structures produced by magnetic buoyancy to be
able to overcome the downward pumping. Alternatively, if $M_{U} \gtrsim 1$, we would expect
the generated field to produce buoyant structures that are too weak to overcome the downward pumping,
unless magnetic flux can be locally concentrated so that $|\mathbf{B}_{h}| >
B_{eq}$. As in the simulations without shear, we study various values
of $M_{U}$ either side of unity by fixing $U_{m}$ and varying
$\gamma_{m}$.

Since the shear is designed to represent the radial shear in the
tachocline, we choose the width of the shear region to be much thinner than a pressure scale
height, with $\Delta z_{s} \approx 0.03 \ll 1$. The magnitude of the
shear is subsonic, with $U_{m} = 0.1$, though this is still larger
than in the Sun so that the evolution can be fully captured within a sensible
run time. Note that this shear is hydrodynamically unstable, with $Ri
\approx 0.074 < 0.25$. This means that the shear excites vortices (aligned in
the $x$-direction) through Kelvin-Helmholtz type instabilities. Since these
instabilities are unwanted, we try to partially reduce their effect by
having a small fluid Reynolds number\footnote{This is defined to be
  $Re = \frac{U_{m}\Delta z}{\sigma C_{K}}$.} for the shear of $Re \sim 66$. In
any case, the resulting vortices are eventually suppressed by the induced horizontal
field.

\section{Results: shear production of a magnetic layer, its
  instability and interaction with downward pumping}
\label{resultswithshear}

In this section we describe the results of a set of simulations with
shear using the numerical model outlined in \S \ref{modelwithshear}. 
First we describe the temporal evolution for our fiducial case, which
has $M_{U}=1$. We then discuss the effects of varying $M_{U}$.

\begin{figure*}
  \begin{center}
\subfigure{\includegraphics[width=0.25\textwidth]{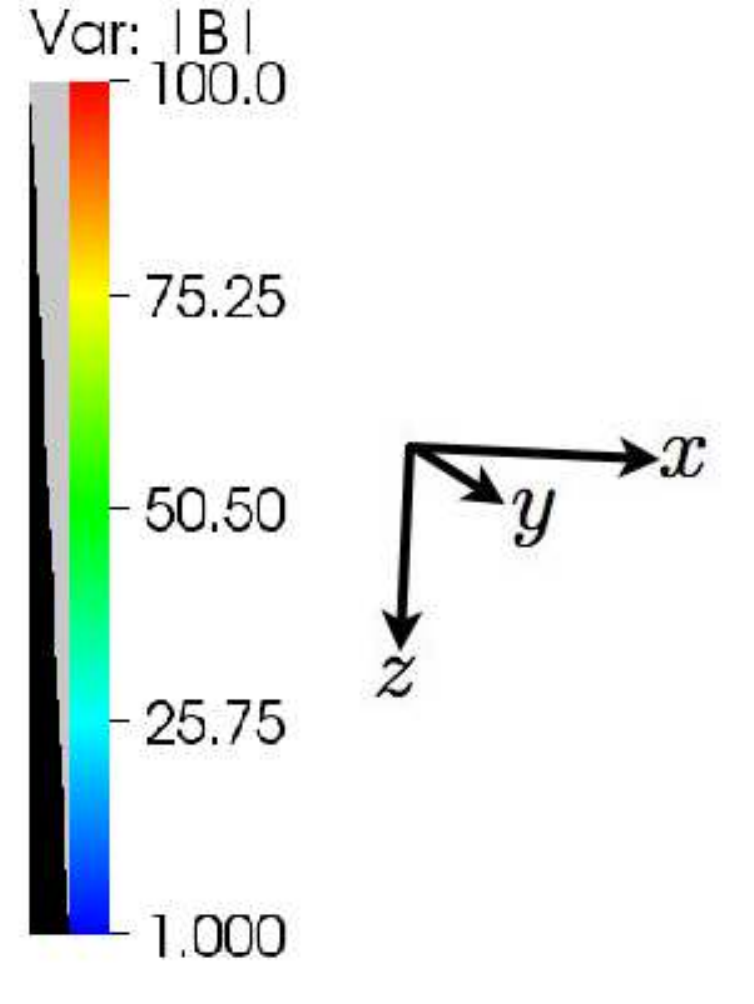} } 
    \subfigure{\includegraphics[width=0.4\textwidth]{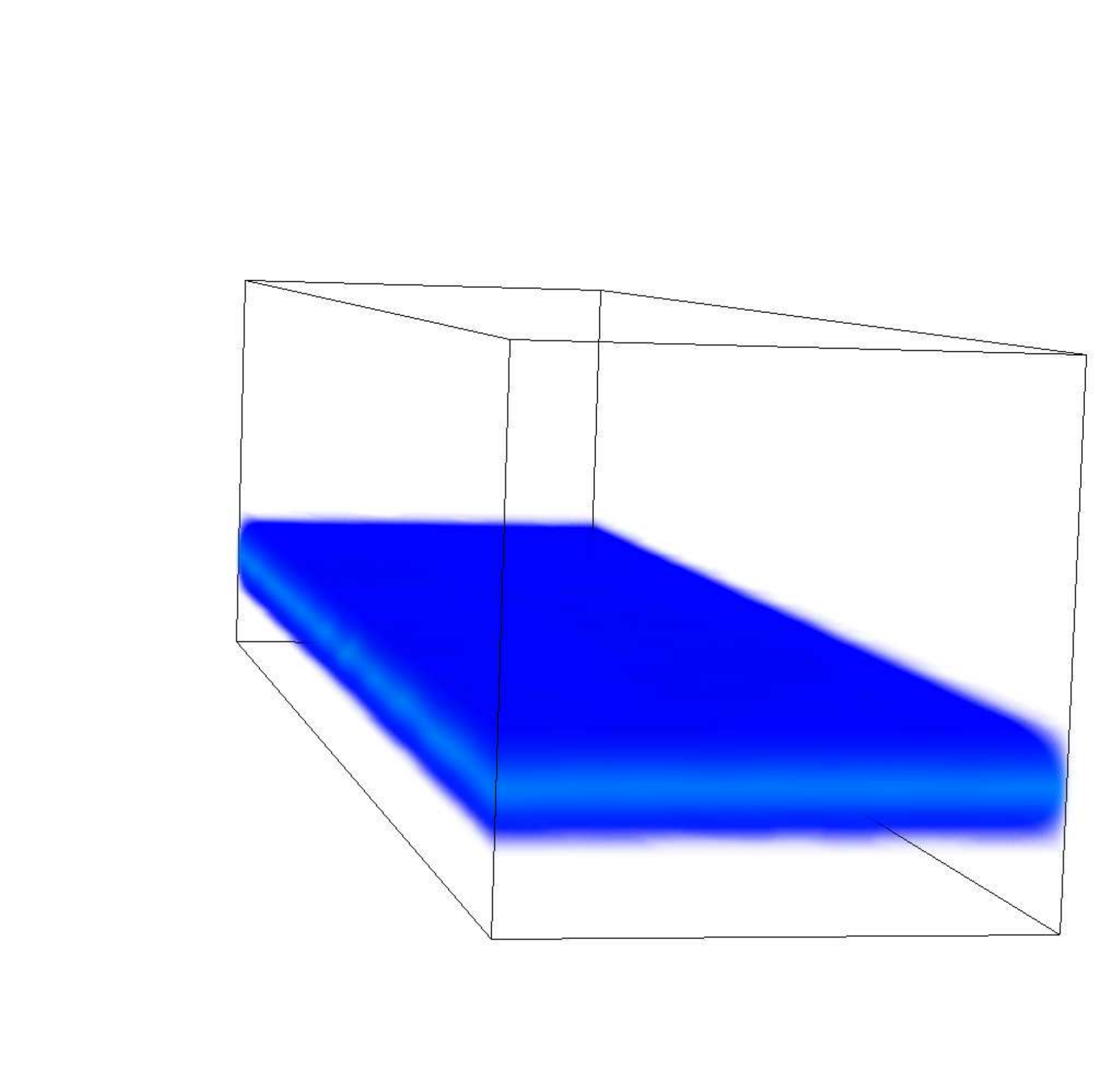} } \\
 \subfigure{\includegraphics[width=0.4\textwidth]{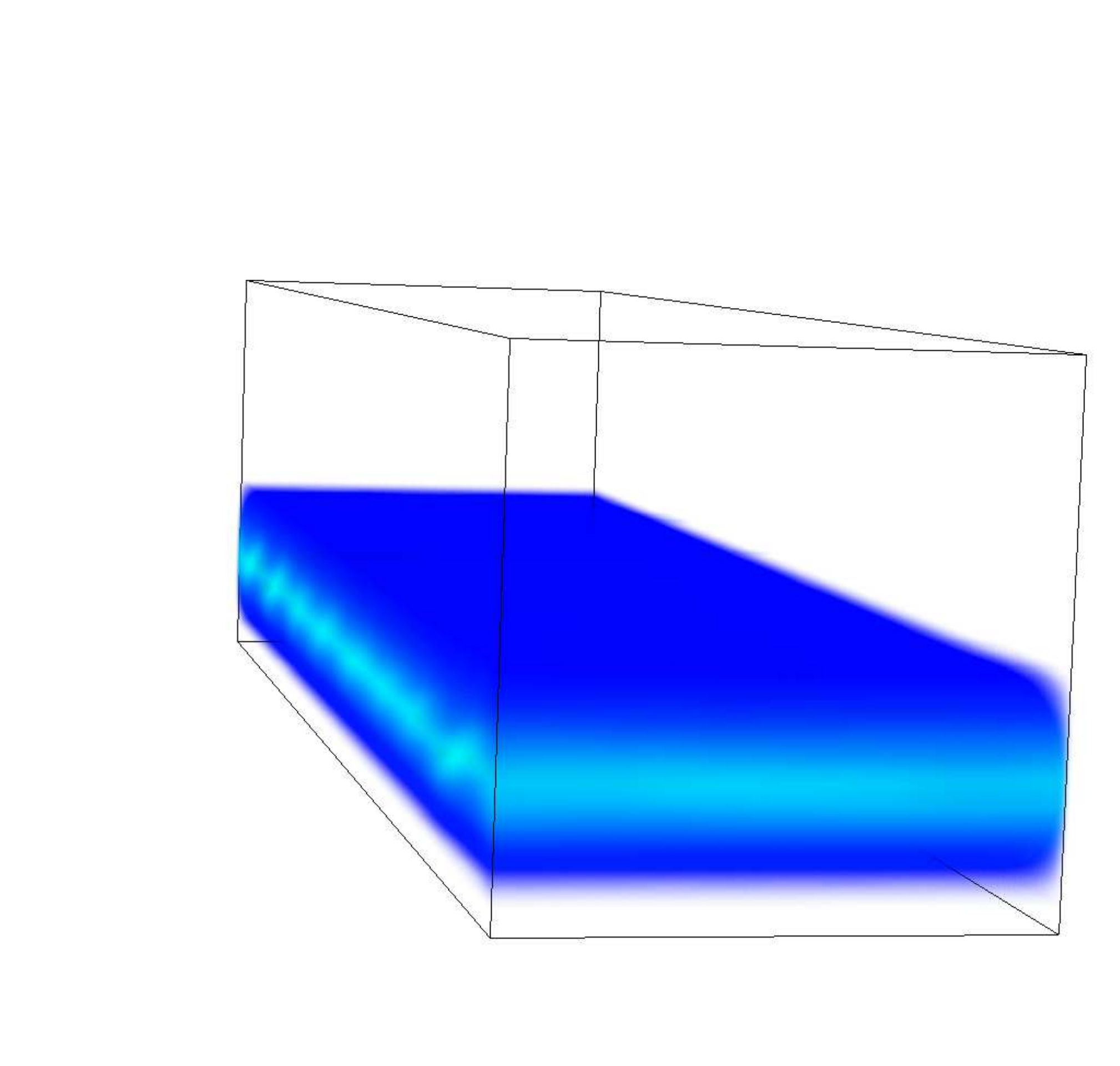} } 
\subfigure{\includegraphics[width=0.4\textwidth]{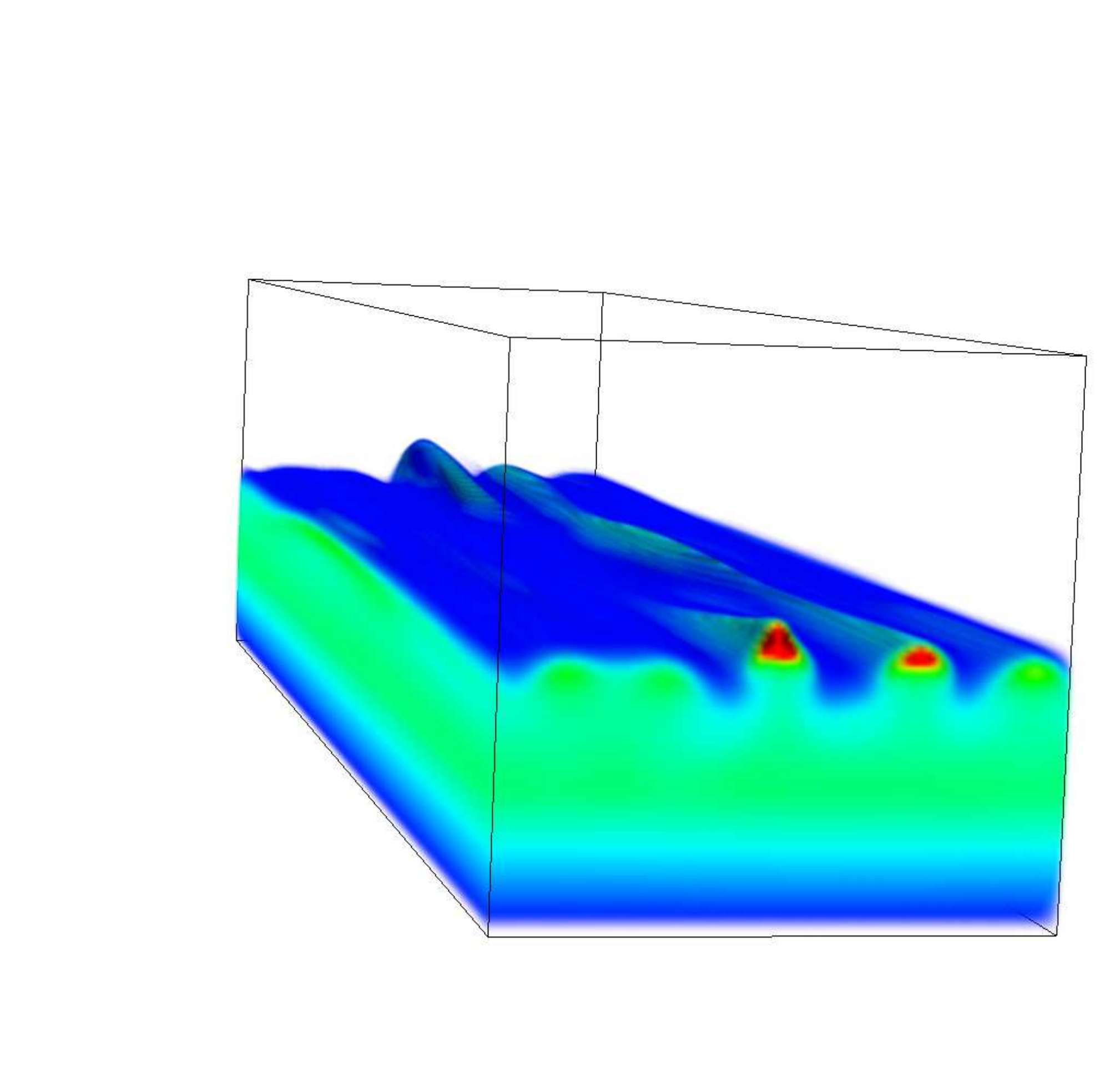} } \\
\subfigure{\includegraphics[width=0.4\textwidth]{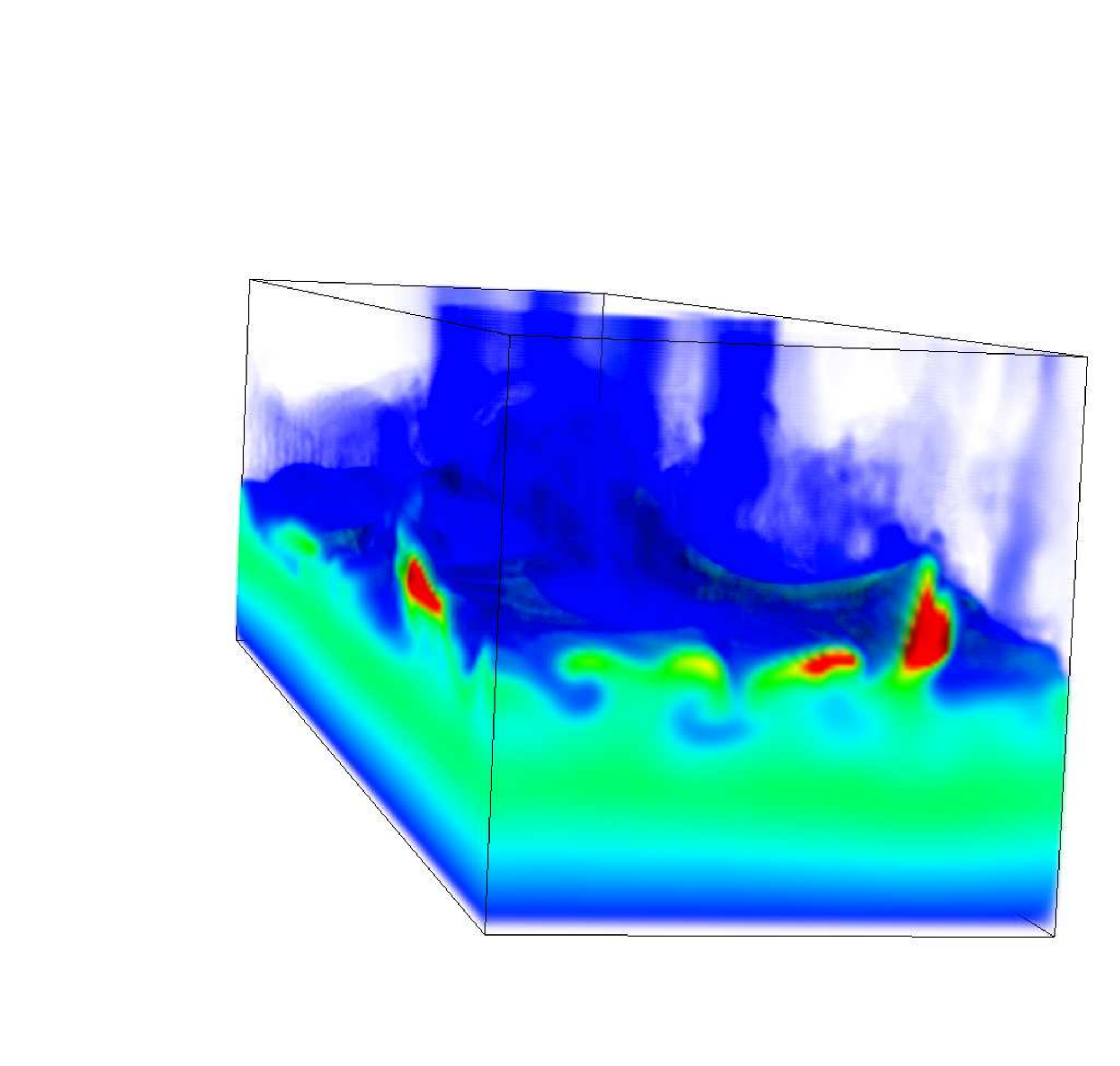} }
\subfigure{\includegraphics[width=0.4\textwidth]{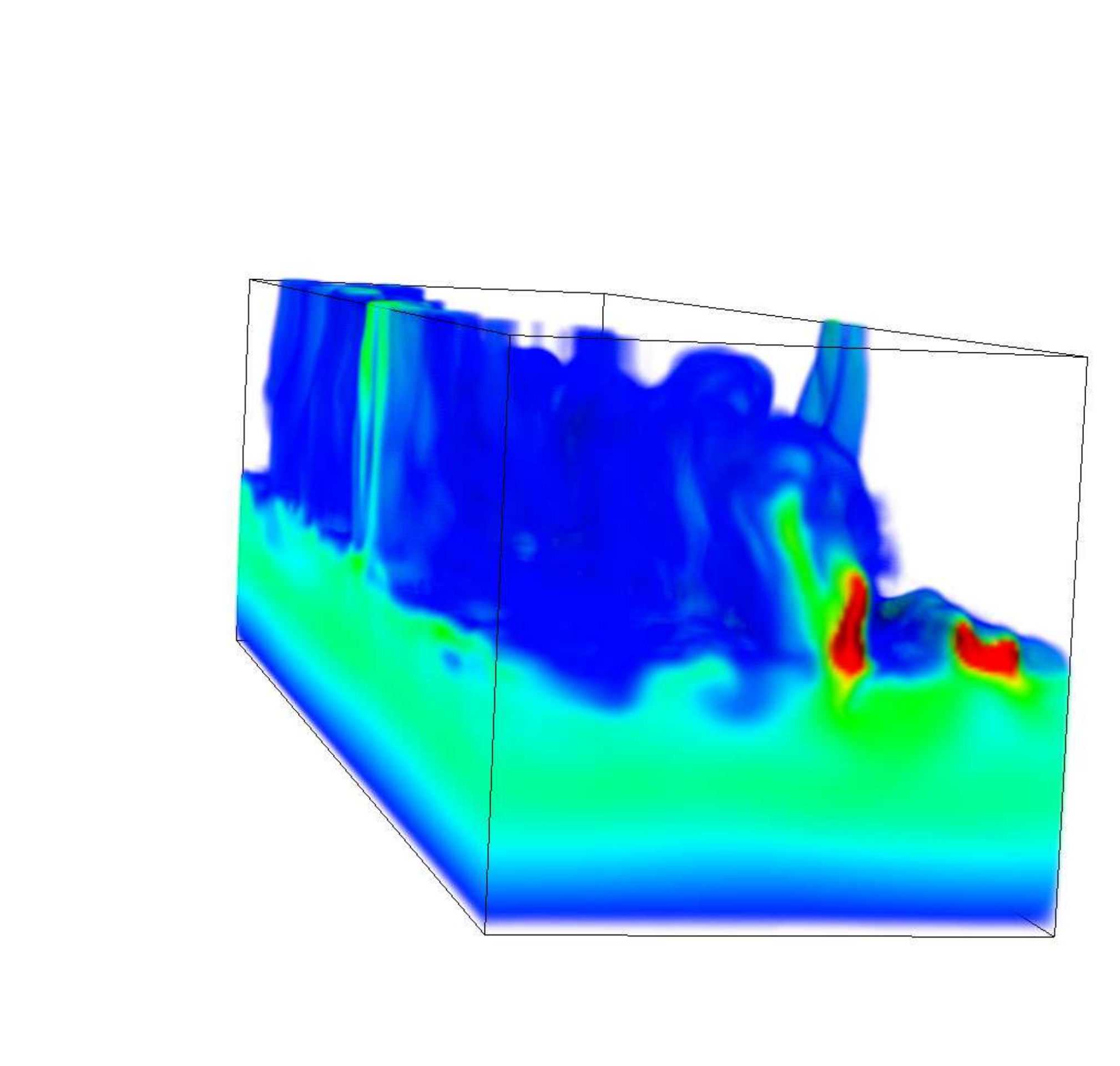} }
  \end{center}
  \caption{Volume renderings of $|\mathbf{B}|$ for a simulation with
    $M_{U}=1$ at approximate times $t = 15, 42, 160, 177$ and $200$, respectively. This illustrates the
    temporal evolution of the magnetic field in a simulation of the
    buoyancy instability of a shear-generated magnetic slab with magnetic
    pumping in the upper layer, using $M_{U}=1$.
 } 
\label{SHKHvolume}
\end{figure*}

As the simulation begins, vertical shear acting on the 
uniform vertical magnetic field generates a magnetic
layer aligned in the (negative) $y$-direction (for positive $U_{m}$).
In the initial stages this field grows in magnitude and also
gradually expands due to the vertical propagation of Alfv\'{e}n waves
along the initial field. A hydrodynamic
Kelvin-Helmholtz type instability sets in from our prescribed shear
flow, which generates vortices aligned in the $x$-direction. This
occurs because our adopted shear is hydrodynamically unstable, as we
explained in \S \ref{modelwithshear}. However, as the magnetic field in the layer grows, Lorentz forces act
back on the resulting flow and
hinder further development of the
instability (e.g.\ SBP09). It therefore eventually dies out as the field
strength exceeds a critical value, leaving behind only some weak inhomogeneities
in the $y$-direction. Volume rendering images illustrating the temporal evolution of our
simulation with $M_{U}=1$ for
$|\mathbf{B}|$ are presented in Fig.~\ref{SHKHvolume}. 

\begin{figure}
  \begin{center}
    \subfigure{\includegraphics[width=0.5\textwidth]{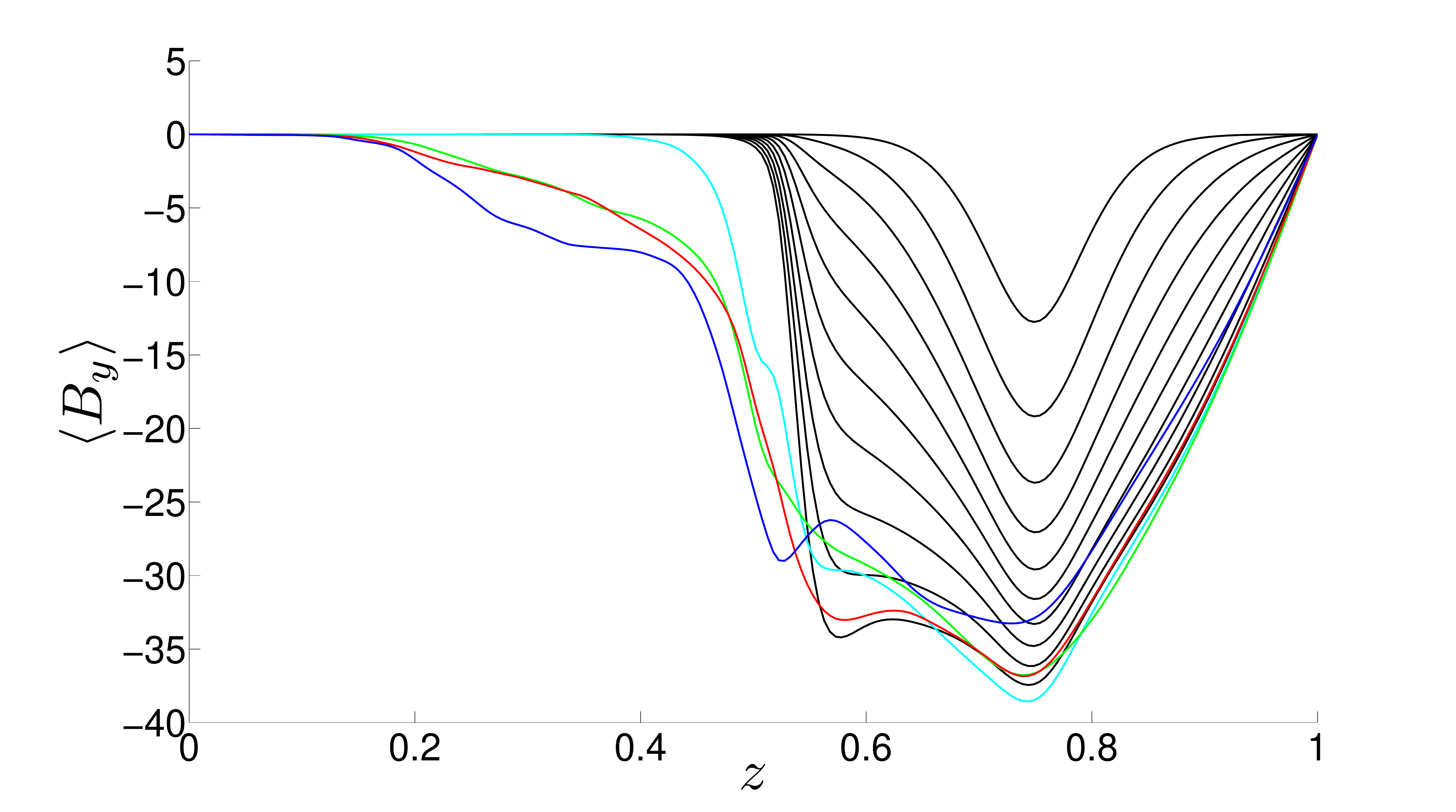} } 
  \end{center}
  \caption{Horizontally integrated $B_{y}$ in a simulation with
    $M_{U}=1$ for multiple
    time intervals between $t\approx 15-150$ until the field has built up a sufficient gradient near
    $z_{i}$ for buoyancy instabilities to set in at this location (black lines),
    together with the same quantity after onset of buoyancy
    instability there, at approximate times $t = 163$ (light blue), $192$
    (green), $222$ (red) and $266$ (dark blue). The maximum
    field amplitude in the shear region is approximately equal to
    $U_{m}/\sqrt{F}$.}
\label{layergrowth}
\end{figure}

Once magnetic field gradients
become sufficiently strong, buoyancy instabilities occur in the upper
parts of this expanding magnetic layer. These are of two-dimensional interchange type (like in
\S~\ref{results1}), which are expected to be least affected by the addition of 
an aligned shear flow \citep{TH2004}.
At the same time, vertically propagating Alfv\'{e}n waves, together with a small amount
of ohmic diffusion, spread the horizontal magnetic
field throughout the lower layer.
Since the field is confined from
above at the $\gamma$-interface, it remains well confined within the lower layer unless its
magnitude can locally exceed $B_{eq}$. For this to occur
once the buoyant magnetic structures, or the bulk of the expanding
magnetic layer, first reach the interface requires $M_{U} \ll 1$,
otherwise the bulk of the field is held down below the
$\gamma$-interface. The maximum field strength in the shear-generated layer reaches
$B_{y} \sim \frac{U_{m}}{\sqrt{F}} \sim 35$. This can be seen in
Fig.~\ref{layergrowth}, which displays the horizontally-integrated
vertical profile of $B_{y}$ as the field builds underneath the
interface for $t=15-150$ (thin black lines). The field for $z \gtrsim
z_{i}$ builds up to approximately this value due to the combined
action of shear at generating the magnetic layer and
$\gamma$-pumping at holding down the field.

There are several important effects resulting from the addition of $\gamma$-pumping in the
upper layer. The simplest of these is to hold down the bulk of the
field in the lower layer. If $\gamma$-pumping is not present, the bulk
of the field, together with any buoyantly unstable pockets of field,
continue to expand throughout the upper layer. Since the Sun has some
means of holding back field until it reaches a certain strength, we
require some mechanism to hold down weak field. In our simulations magnetic flux pumping
in the upper layer (``the convection zone'') provides such a
mechanism, with only localised pockets of strong
field able to rise above $z\approx z_{i}$. Unlike in \S \ref{results1}, where the
interactions between vortices are also capable of holding down the
bulk of the magnetic field, in these simulations the magnetic layer
would continue to expand because of the slow vertical propagation of Alfv\'{e}n
waves along the initial field. We therefore require $\gamma$-pumping
to hold down the bulk of the field in these simulations.

Another important effect of $\gamma$-pumping is to produce strong vertical magnetic field
gradients in the vicinity of $z_{i}$ (see Fig.~\ref{layergrowth}), which induces buoyancy
instabilities at this location. These result from the 
vertical gradients of $\gamma$, and provide an additional way to
produce strong field gradients, and therefore induce (in our case non-diffusive) buoyancy
instabilities, without requiring strong (i.e.\ hydrodynamically
unstable) shear (e.g.\ VB08). Once the upper interface of
the magnetic layer is perturbed by the instability, the important
factor is the relative
strength of the field in the rising magnetic structures to that which can be held down by the
$\gamma$-pumping, i.e.\ whether $|\mathbf{B}_{h}|>B_{eq}$.

As in \S \ref{results1}, rising pockets of magnetic field generate
vortices, so the vicinity of $z_{i}$ is subject to complicated 
interactions between them. 
In some cases, the resulting fluid motions are able to concentrate the
field horizontally below the interface into localised pockets of strong field (an
example of this is plotted later in Fig.~\ref{SHKHMU2volume}). Note
that the spatial scales of these magnetic structures are not necessarily the
same as those of the initial buoyancy instability. A combination of the above effects,
together with the continual forcing of the horizontal layer by the
shear, can produce localised peak fields that satisfy $|\mathbf{B}_{h}|>B_{eq}$. Such
strong pockets can rise into the upper layer. As before, rising flux
structures are eventually sheared apart by shear
interactions with the gas in the upper layer, and do not rise far as
coherent structures.

 \begin{figure}
   \begin{center}
     \subfigure{\includegraphics[width=0.5\textwidth]{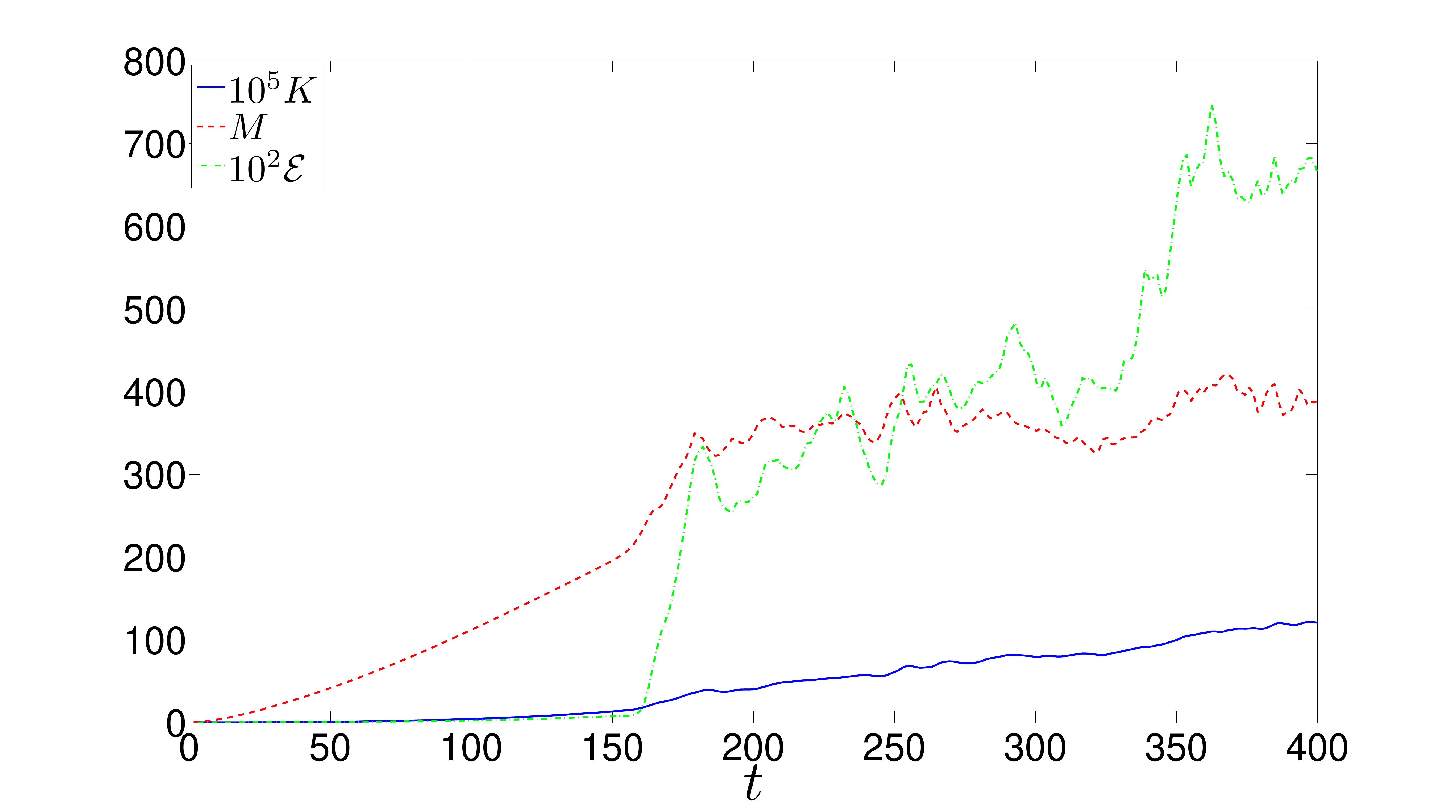} }
   \end{center}
   \caption{Temporal evolution of the 
     volume-integrated magnetic energy $M$, kinetic energy $K$ and
     enstrophy $\mathcal{E}$, in a simulation with shear and
     $M_{U}=1$, where these quantities have been scaled as listed in
     the legend.} 
 \label{SHEnergy}
 \end{figure}

As is illustrated in Fig.~\ref{SHEnergy}, the magnetic energy builds until it reaches its
peak value just before the field below the interface has been
sufficiently concentrated to be able to break out into
the upper layer. Once buoyancy instabilities occur, the resulting
local shear excites vortices, thereby increasing $K$ and
$\mathcal{E}$. Since we are constantly forcing the system through the
steady shear flow $\mathbf{U}_{0}$, the magnetic energy oscillates
about an approximately constant value, and does not appreciably
decay throughout the simulation. We analyse the results until
$t\approx 400$, after which the influence of the upper boundary becomes
important.

\begin{figure}
  \begin{center}
    \subfigure{\includegraphics[width=0.5\textwidth]{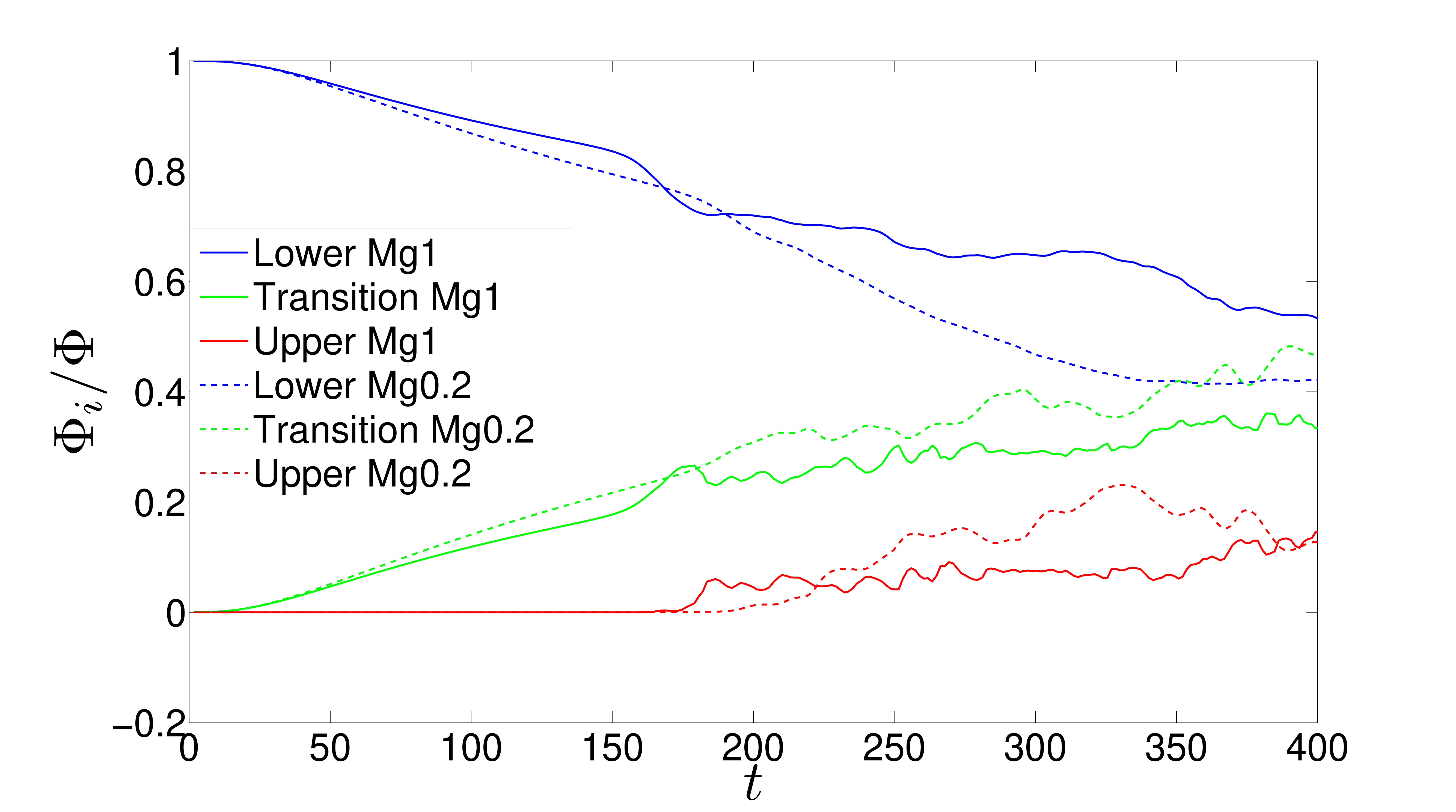} } 
  \end{center}
  \caption{Temporal evolution
    of the magnetic flux fraction contained within the lower layer $\Phi_{L}/\Phi$, the
    transition region $\Phi_{T}/\Phi$ and the upper layer
    $\Phi_{U}/\Phi$, in a simulation with $M_{\gamma}=1$ (solid lines)
    and $M_{\gamma}=0.2$ (dashed lines).} 
\label{SHflux}
\end{figure}

\begin{figure}
  \begin{center}
    \subfigure{\includegraphics[width=0.5\textwidth]{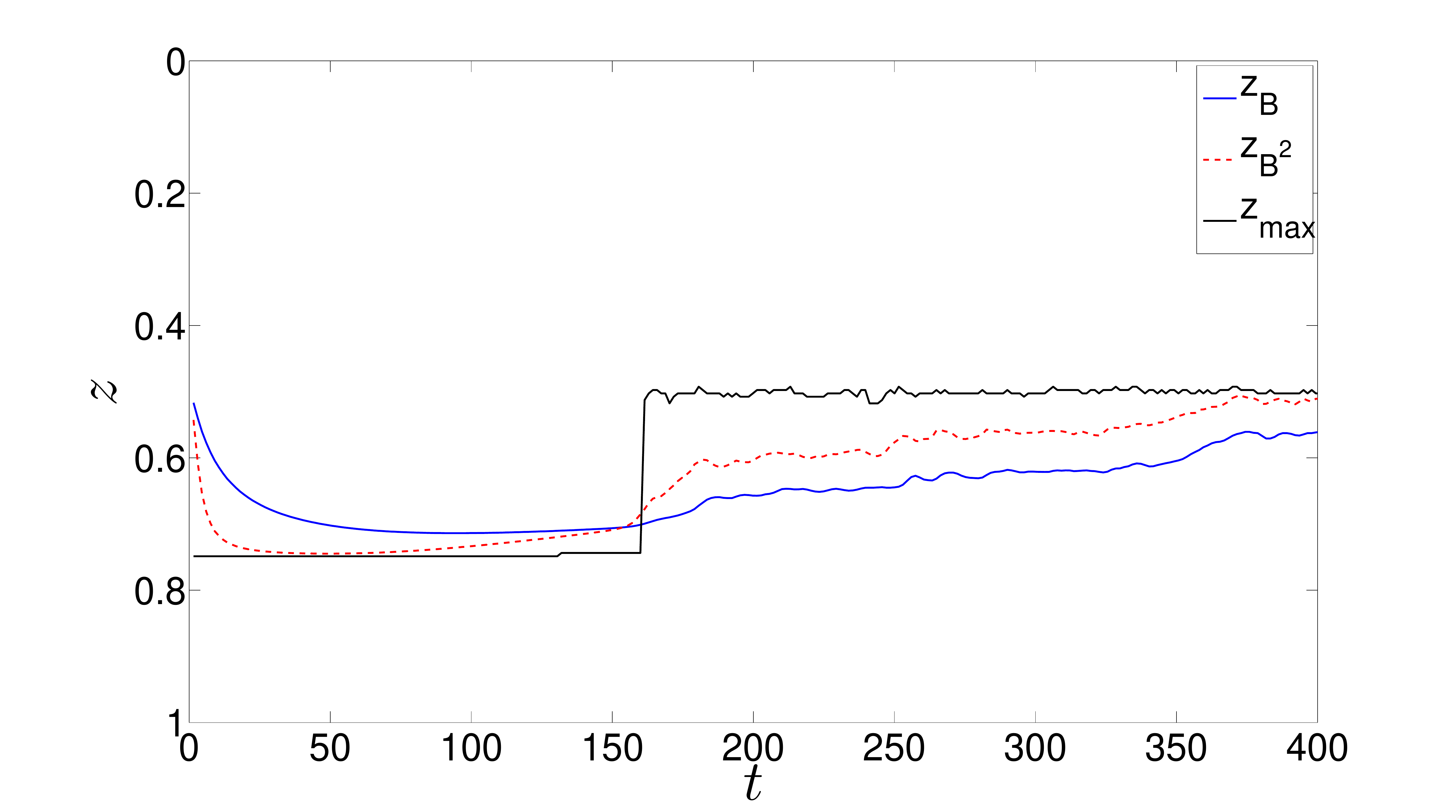} }
  \end{center}
  \caption{Temporal evolution of the peak magnetic field $z_{\mathrm{max}}$,
    the centre of magnetic field $z_{B}$, and the centre of
    magnetic energy $z_{B^{2}}$ for a simulation with $M_{U}=1$.} 
\label{SHCentrefield}
\end{figure}

We plot the flux fractions contained in each layer normalised to the
total $\Phi$ (which is a time varying quantity due to shear production
and ohmic dissipation) at each time in
Fig.~\ref{SHflux}. This figure shows that the magnetic flux is
initially produced and contained within the lower layer until the
field has expanded and reached $z\approx z_{i}$. Afterwards,
magnetic flux is primarily stored in the lower layer and transition
region, with only a few localised outbursts transporting flux into the
upper layer. We stress the localised nature of these pockets of field
that rise into the upper layer, since the bulk of the field is well
confined within the lower layer (i.e.\ $|\Phi_{U}| \ll |\Phi|$ at all
times). We also plot the various measures of the distribution of
magnetic field defined by Eqs.\ \ref{measure1}--\ref{measure3} as a
function of time in Fig.~\ref{SHCentrefield}, where it can be seen
that they are each located within the transition region.  This is the
case even when strong localised breakouts are protruding flux into the
upper layer. Note that again $z_{B^{2}}$ is located slightly higher
than $z_{B}$. The peak magnetic field always remains at $z\approx
z_{i}$.

\subsection{Variation of $M_{U}$}

\begin{figure*}
  \begin{center}
 \subfigure{\includegraphics[width=0.49\textwidth]{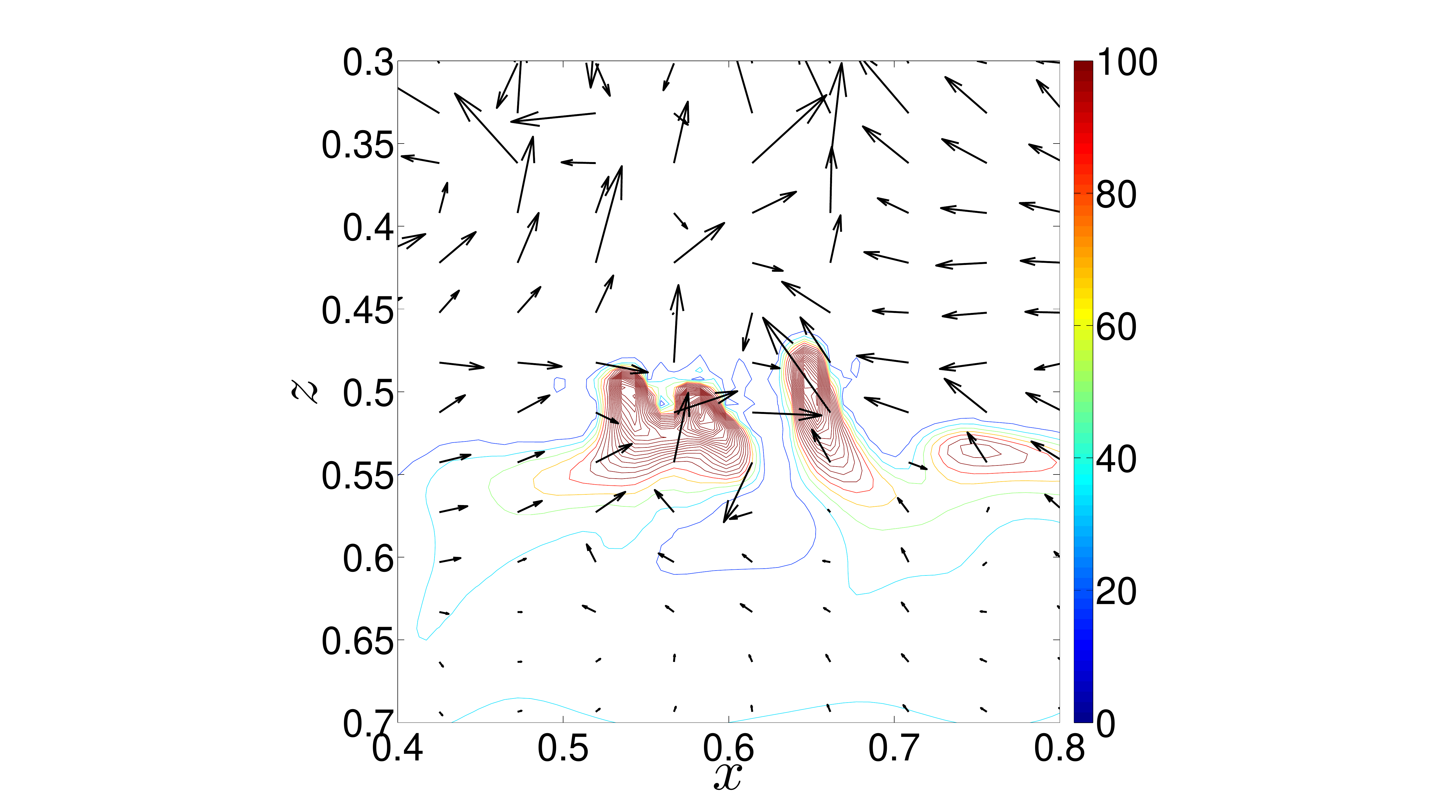} } 
 \subfigure{\includegraphics[width=0.49\textwidth]{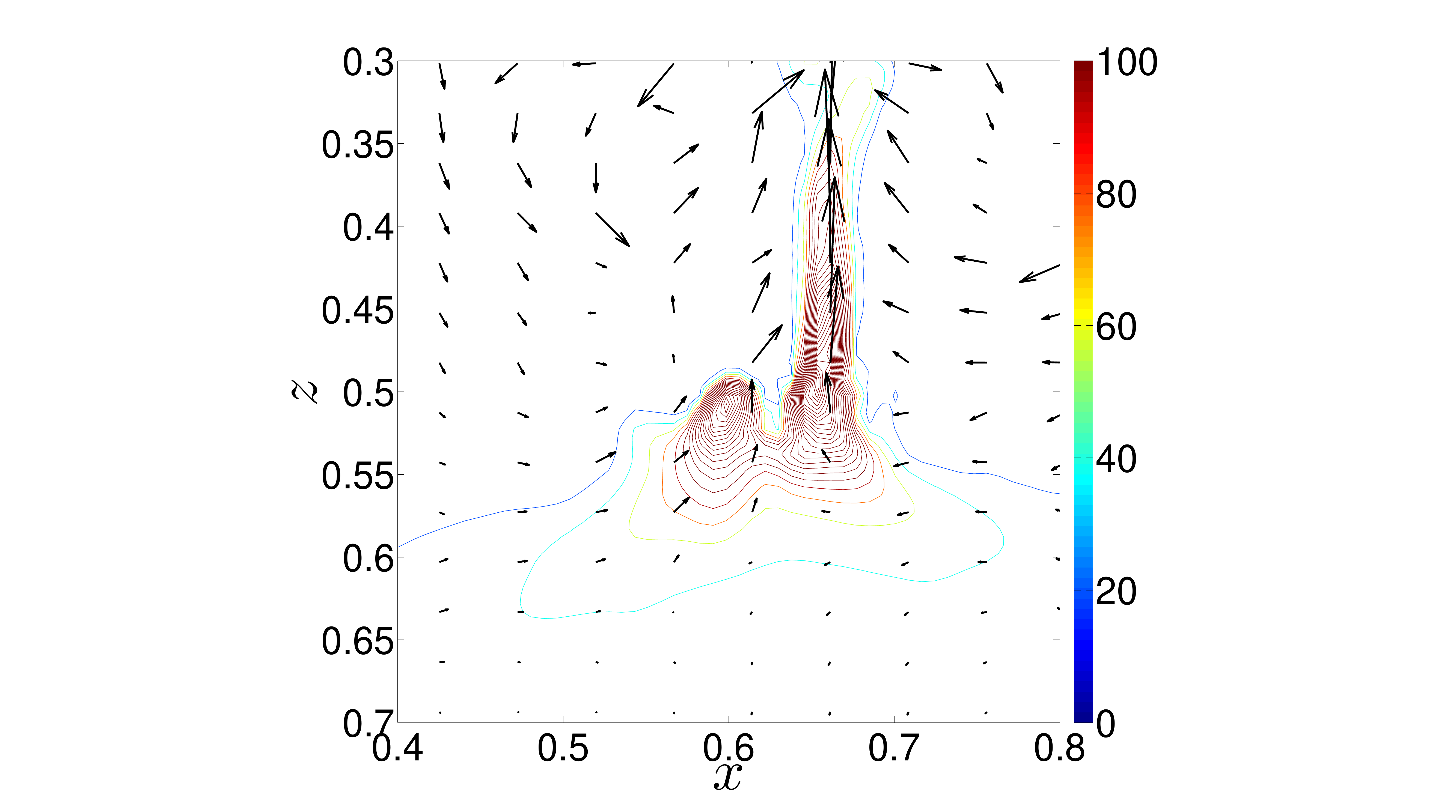} } \\
  \end{center}
  \caption{Contour plot of $|\mathbf{B}|$, together with velocity
    vectors at equally spaced points in a region of the $xz$-plane, in a simulation with
    $M_{U}=2$ at approximate times $t=166$ (left) and $169$ (right). This
    illustrates the concentration of magnetic flux in the transition
    region by vortical fluid motions, which can produce localised pockets of
    strong magnetic field with strengths sufficient to rise into the upper layer.
 } 
\label{SHKHMU2volume}
\end{figure*}

We have varied $M_{U}$ and looked at various values either side of unity to
study the evolution in cases where the kinetic energy of the shear
does and does not exceed that associated with the downward pumping. In the
$M_{U}>1$ regime, we might expect the shear-generated field to be unable to reach strengths sufficient for
buoyant rise into the upper layer. However, this neglects the
concentration of horizontal magnetic flux by vortical
fluid motions, produced in the nonlinear stages of evolution of the system.
We have observed the field to be locally concentrated and to rise into the upper layer in localised
breakouts even when $M_{U}=4$, i.e.\ the kinetic energy of the shear is
significantly less than $(1/2)\rho\gamma_{m}^{2}$, and therefore
also smaller than the kinetic energy of the convection, by more than an order
of magnitude. In Fig.~\ref{SHKHMU2volume}, we show an example of the concentration of
field in the transition region by vortical fluid motions in the
vicinity of $z_{i}$ in a simulation with $M_{U}=2$. A localised pocket of field with a
strength $\mathbf{B}_{h} \gtrsim 100 \sim B_{eq}$ is produced, which then breaks out into the upper layer.

Note that in these simulations
$\gamma_{m}$ is not as strongly subsonic as it is likely to be in
the tachocline. The plasma $\beta$ required for a pocket of magnetic
field to rise is therefore much smaller than we expect in
reality, and for the strongest field $\beta \ll 1$. The magnetic pressure is therefore able to evacuate the gas
within a grid cell, and this becomes particularly important in the
strongest field pockets produced when
$M_{U} > 1$. This causes the timestep to decrease towards zero to satisfy the
Courant-Friedrichs-Lewy stability constraint, and the numerical code
to fail. This limits the maximum value of $M_{U}$ that we can
simulate to $M_{U} \lesssim 5$, unless we either increase the
resolution or reduce the initial values of $\gamma_{m}$ and $U_{m}$,
which both require greater computational resources. Nevertheless,
simulations with $1 \lesssim M_{U} \lesssim 5$ indicate that it is
possible in this regime for a combination of $\gamma$-pumping at
holding down (and amplifying) the field, and vortical fluid motions at concentrating
magnetic flux, to produce pockets with sufficient field strength to be able to rise into
the upper layer.

\section{Conclusions}
\label{discussion}

In this paper we have studied the interaction between magnetic
buoyancy instabilities and magnetic flux pumping in simplified
numerical models of the solar
tachocline. We have adopted an idealised ``mean-field'' model of magnetic
flux pumping, in which a spatially uniform and temporally constant downward
advective velocity $\gamma$ for the magnetic field is added into the induction
equation in an upper layer, with no flux pumping present in the lower
layer. This situation is designed to crudely represent the effects of magnetic
flux pumping in the lower parts of the convection zone, which overly a stable
radiative region containing a layer of buoyantly unstable toroidal magnetic field.

We first studied the addition of a simple $\gamma$-effect into
simulations of the instability of a preconceived horizontal slab of
magnetic field, extending the calculations of \cite{CH1988} and
\cite*{Matthews1995b}. In this problem, the initial configuration is a
magnetostatic equilibrium, perturbed by small thermal perturbations,
which induce Rayleigh-Taylor type instabilities at the top of the magnetic
layer. The resulting magnetic mushrooms rise until they reach the
pumping layer. The effect of this layer on the resulting evolution
depends on the ratio of the downward pumping velocity to the
Alfv\'{e}n speed of the magnetic field, that we denote $M_{\gamma}$. 
When $M_{\gamma}\gtrsim 1$, the magnetic flux pumping
effectively holds down the bulk of the field, only allowing localised pockets of
strong field to rise, which have been
concentrated by vortical fluid motions.

Since the Rayleigh-Taylor type instabilities of a toroidal magnetic
layer do not have a weak field cut off, and occur for any field
strength (in the absence of diffusion), there must exist a mechanism
to hold down the field until it can exceed a critical strength \citep{Hughes2007}. Magnetic
flux pumping can provide one solution to this problem, since its
addition immediately prevents fields weaker than equipartition
strength from rising into the convection zone. If $M_{\gamma} \gtrsim 1$ in reality, then
magnetic flux pumping can explain why the bulk of the field is stored
in the radiative interior, with only localised pockets of strong field
able to rise through the convection zone towards the surface. It is
interesting to note that our simulations in this regime show that the instability of
a \textit{uniform} initial field lying underneath a
layer with \textit{uniform} downward magnetic flux pumping can produce
\textit{localised} clumps of field that rise some distance into the upper layer.

We also studied the addition of a $\gamma$-effect into simulations
of the instability of a shear-generated magnetic layer, continuing
calculations in the spirit of VB08 and SBP09. In this problem, we consider radial
tachocline shear to induce a toroidal magnetic layer, which then becomes buoyantly unstable. The effect of magnetic flux pumping on
this problem has several important contributions. One is to produce
strong magnetic field gradients near the interface of the pumping
layer. This strongly enhances the likelihood of buoyancy instabilities
occurring in this region, and is therefore one method of inducing such
instabilities when the toroidal magnetic layer is forced by a weak, 
hydrodynamically stable, tachocline shear. We do not, therefore, 
require strong shear for magnetic buoyancy instabilities to be
excited (c.f.\ VB08; VB09).

The evolution in cases with shear depends on the ratio of the pumping
velocity to the shear velocity, that we denote $M_{U}$. 
One interesting result of these simulations is that even in the case
in which $M_{U} > 1$ the shear was able to produce localised rising pockets
of field. This is interesting because the tachocline shear is probably maintained
at a level such that it has a similar mean kinetic energy density as
the convection. Therefore, the toroidal field that is produced by this
shear might be expected to have at most equipartition strength with the convective
downflows. Since we have observed localised pockets of magnetic flux to rise in the regime
with $M_{U} > 1$, because field is amplified by the combined action of
concentration by vortical fluid motions,
shear at forcing the layer and $\gamma$-pumping at holding it down
(and amplifying it through its non-zero divergence),
this indicates that the shear is not necessarily required to
be more energetic than the convection for superequipartition fields to
be produced by magnetic buoyancy instabilities. 
Somewhat paradoxically, magnetic flux pumping (or
  rather its strong radial gradients near the tachocline) may in fact be an
essential ingredient in producing localised pockets of superequipartition field that are
able to rise up through the convection zone and to the solar surface.

One important problem, which we have not attempted to address here, is how does the
instability produce fields that are sufficiently helical for the
resulting magnetic structures to survive their passage through the
convection zone? The solution to this problem will require
consideration of initial fields that are spatially inhomogeneous and
not unidirectional. In addition, we have modelled the effects of
magnetic flux pumping in the crudest possible way. Indeed, this calculation does not in any sense aim to be the last word on the matter. Rather, it should be seen as a pilot project which has the limited aim of establishing the efficacy of a mechanism for flux concentration. Now that this mechanism has been shown to have validity the next step is a fully resolved calculation for more general initial fields and a turbulent convection zone in which the effect can be put on a proper quantitative footing. This work is presently in progress.

\section*{Acknowledgements}

We would like to thank Laur\`{e}ne Jouve and William Edmunds for
useful discussions during the early stages of this work, and the anonymous referee for carefully reading the manuscript. This work was funded by an STFC rolling grant. Some of the simulations were performed using the High Performance Computing Service at the University of
Cambridge.

\setlength{\bibsep}{0pt}
\bibliography{mbbib}

\begin{thebibliography}{}

\bibitem[\protect\citeauthoryear{{Acheson}}{{Acheson}}{1979}]{Acheson1979}
{Acheson} D.~J.,  1979, Solar Physics, 62, 23

\bibitem[\protect\citeauthoryear{{Arter}}{{Arter}}{1983}]{Arter1983}
{Arter} W.,  1983, Journal of Fluid Mechanics, 132, 25

\bibitem[\protect\citeauthoryear{{Arter}, {Proctor} \& {Galloway}}{{Arter}
  et~al.}{1982}]{Arteretal1982}
{Arter} W.,  {Proctor} M.~R.~E.,    {Galloway} D.~J.,  1982, MNRAS, 201, 57P

\bibitem[\protect\citeauthoryear{{Brummell}, {Cline} \& {Cattaneo}}{{Brummell}
  et~al.}{2002}]{BCC2002}
{Brummell} N.,  {Cline} K.,    {Cattaneo} F.,  2002, MNRAS, 329, L73

\bibitem[\protect\citeauthoryear{{Bushby} \& {Houghton}}{{Bushby} \&
  {Houghton}}{2005}]{Bushby2005}
{Bushby} P.~J.,  {Houghton} S.~M.,  2005, MNRAS, 362, 313

\bibitem[\protect\citeauthoryear{{Cattaneo} \& {Hughes}}{{Cattaneo} \&
  {Hughes}}{1988}]{CH1988}
{Cattaneo} F.,  {Hughes} D.~W.,  1988, Journal of Fluid Mechanics, 196, 323

\bibitem[\protect\citeauthoryear{{Cattaneo}, {Hughes} \& {Proctor}}{{Cattaneo}
  et~al.}{1988}]{CHP88}
{Cattaneo} F.,  {Hughes} D.~W.,    {Proctor} M.~R.~E.,  1988, Geophysical and
  Astrophysical Fluid Dynamics, 41, 335

\bibitem[\protect\citeauthoryear{{Christensen-Dalsgaard} \&
  {Thompson}}{{Christensen-Dalsgaard} \& {Thompson}}{2007}]{JCDMT2007}
{Christensen-Dalsgaard} J.,  {Thompson} M.~J.,  2007, in {D.~W.~Hughes,
  R.~Rosner, \& N.~O.~Weiss} ed., The Solar Tachocline {Observational results
  and issues concerning the tachocline}.
p.~53

\bibitem[\protect\citeauthoryear{{Cline}, {Brummell} \& {Cattaneo}}{{Cline}
  et~al.}{2003}]{CBCdynamo2003}
{Cline} K.~S.,  {Brummell} N.~H.,    {Cattaneo} F.,  2003, ApJ, 599, 1449

\bibitem[\protect\citeauthoryear{{Crow}}{{Crow}}{1970}]{Crow1970}
{Crow} S.~C.,  1970, AIAA Journal, 8, 2172

\bibitem[\protect\citeauthoryear{{Dorch} \& {Nordlund}}{{Dorch} \&
  {Nordlund}}{2001}]{DorchNordlund2001}
{Dorch} S.~B.~F.,  {Nordlund} {\AA}.,  2001, A\& A, 365, 562

\bibitem[\protect\citeauthoryear{{Drobyshevski} \& {Yuferev}}{{Drobyshevski} \&
  {Yuferev}}{1974}]{DrobYuf1974}
{Drobyshevski} E.~M.,  {Yuferev} V.~S.,  1974, Journal of Fluid Mechanics, 65,
  33

\bibitem[\protect\citeauthoryear{{Galloway} \& {Proctor}}{{Galloway} \&
  {Proctor}}{1983}]{GallProc83}
{Galloway} D.~J.,  {Proctor} M.~R.~E.,  1983, Geophysical and Astrophysical
  Fluid Dynamics, 24, 109

\bibitem[\protect\citeauthoryear{{Gilman}}{{Gilman}}{1970}]{Gilman1970}
{Gilman} P.~A.,  1970, ApJ, 162, 1019

\bibitem[\protect\citeauthoryear{{Gough}}{{Gough}}{2007}]{Gough2007}
{Gough} D.,  2007, in {D.~W.~Hughes, R.~Rosner, \& N.~O.~Weiss} ed., The Solar
  Tachocline {An introduction to the solar tachocline}.
p.~3

\bibitem[\protect\citeauthoryear{{Guerrero} \& {K{\"a}pyl{\"a}}}{{Guerrero} \&
  {K{\"a}pyl{\"a}}}{2011}]{Kapyla2011}
{Guerrero} G.,  {K{\"a}pyl{\"a}} P.,  2011, ArXiv e-prints

\bibitem[\protect\citeauthoryear{{Hughes}}{{Hughes}}{2007}]{Hughes2007}
{Hughes} D.~W.,  2007, in {D.~W.~Hughes, R.~Rosner, \& N.~O.~Weiss} ed., The
  Solar Tachocline {Magnetic buoyancy instabilities in the tachocline}.
p.~275

\bibitem[\protect\citeauthoryear{{Hughes} \& {Falle}}{{Hughes} \&
  {Falle}}{1998}]{FalleHughes1998b}
{Hughes} D.~W.,  {Falle} S.~A.~E.~G.,  1998, ApJL, 509, L57

\bibitem[\protect\citeauthoryear{{Hughes}, {Falle} \& {Joarder}}{{Hughes}
  et~al.}{1998}]{FalleHughes1998a}
{Hughes} D.~W.,  {Falle} S.~A.~E.~G.,    {Joarder} P.,  1998, MNRAS, 298, 433

\bibitem[\protect\citeauthoryear{{Hughes} \& {Proctor}}{{Hughes} \&
  {Proctor}}{1988}]{HughesProctorReview1988}
{Hughes} D.~W.,  {Proctor} M.~R.~E.,  1988, Annual Review of Fluid Mechanics,
  20, 187

\bibitem[\protect\citeauthoryear{{Jones}, {Thompson} \& {Tobias}}{{Jones}
  et~al.}{2010}]{Jones2010}
{Jones} C.~A.,  {Thompson} M.~J.,    {Tobias} S.~M.,  2010, Space Science
  Reviews, 152, 591

\bibitem[\protect\citeauthoryear{{Jouve} \& {Brun}}{{Jouve} \&
  {Brun}}{2009}]{JouveBrun2009}
{Jouve} L.,  {Brun} A.~S.,  2009, ApJ, 701, 1300

\bibitem[\protect\citeauthoryear{{Kitchatinov} \& {R{\"u}diger}}{{Kitchatinov}
  \& {R{\"u}diger}}{2008}]{KitRud2008}
{Kitchatinov} L.~L.,  {R{\"u}diger} G.,  2008, Astronomische Nachrichten, 329,
  372

\bibitem[\protect\citeauthoryear{{MacGregor} \& {Charbonneau}}{{MacGregor} \&
  {Charbonneau}}{1999}]{MacGregor1999}
{MacGregor} K.~B.,  {Charbonneau} P.,  1999, ApJ, 519, 911

\bibitem[\protect\citeauthoryear{{Matthews}, {Hughes} \& {Proctor}}{{Matthews}
  et~al.}{1995}]{Matthews1995b}
{Matthews} P.~C.,  {Hughes} D.~W.,    {Proctor} M.~R.~E.,  1995, ApJ, 448, 938

\bibitem[\protect\citeauthoryear{{Matthews}, {Proctor} \& {Weiss}}{{Matthews}
  et~al.}{1995}]{Matthews1995a}
{Matthews} P.~C.,  {Proctor} M.~R.~E.,    {Weiss} N.~O.,  1995, Journal of
  Fluid Mechanics, 305, 281

\bibitem[\protect\citeauthoryear{{Miesch}, {Brun}, {De Rosa} \&
  {Toomre}}{{Miesch} et~al.}{2008}]{Miesch2008}
{Miesch} M.~S.,  {Brun} A.~S.,  {De Rosa} M.~L.,    {Toomre} J.,  2008, ApJ,
  673, 557

\bibitem[\protect\citeauthoryear{{Moffatt}}{{Moffatt}}{1978}]{Moffatt1978}
{Moffatt} H.~K.,  1978, {Magnetic field generation in electrically conducting
  fluids}

\bibitem[\protect\citeauthoryear{{Moffatt}}{{Moffatt}}{1983}]{Moffatt1983}
{Moffatt} H.~K.,  1983, Reports on Progress in Physics, 46, 621

\bibitem[\protect\citeauthoryear{{Ossendrijver}}{{Ossendrijver}}{2003}]{Ossend%
rijver2003}
{Ossendrijver} M.,  2003, A\&A Review, 11, 287

\bibitem[\protect\citeauthoryear{{Ossendrijver}, {Stix}, {Brandenburg} \&
  {R{\"u}diger}}{{Ossendrijver} et~al.}{2002}]{Ossendrijver2002}
{Ossendrijver} M.,  {Stix} M.,  {Brandenburg} A.,    {R{\"u}diger} G.,  2002,
  A\& A, 394, 735

\bibitem[\protect\citeauthoryear{{Parker}}{{Parker}}{1975}]{Parker1975}
{Parker} E.~N.,  1975, ApJ, 198, 205

\bibitem[\protect\citeauthoryear{{R{\"a}dler}}{{R{\"a}dler}}{1968}]{Radler1968}
{R{\"a}dler} K.~H.,  1968, Zeitschrift Naturforschung Teil A, 23, 1851

\bibitem[\protect\citeauthoryear{{Rogers}}{{Rogers}}{2011}]{Rogers2011}
{Rogers} T.~M.,  2011, ApJ, 733, 12

\bibitem[\protect\citeauthoryear{{Schmitt} \& {Rosner}}{{Schmitt} \&
  {Rosner}}{1983}]{SR1983}
{Schmitt} J.~H.~M.~M.,  {Rosner} R.,  1983, ApJ, 265, 901

\bibitem[\protect\citeauthoryear{{Silvers}, {Bushby} \& {Proctor}}{{Silvers}
  et~al.}{2009}]{SBP2009}
{Silvers} L.~J.,  {Bushby} P.~J.,    {Proctor} M.~R.~E.,  2009, MNRAS, 400, 337

\bibitem[\protect\citeauthoryear{{Silvers}, {Vasil}, {Brummell} \&
  {Proctor}}{{Silvers} et~al.}{2009}]{SVBP2009}
{Silvers} L.~J.,  {Vasil} G.~M.,  {Brummell} N.~H.,    {Proctor} M.~R.~E.,
  2009, ApJL, 702, L14

\bibitem[\protect\citeauthoryear{{Silvers}, {Vasil}, {Brummell} \&
  {Proctor}}{{Silvers} et~al.}{2010}]{Lara2011}
{Silvers} L.~J.,  {Vasil} G.~M.,  {Brummell} N.~H.,    {Proctor} M.~R.~E.,
  2010, Proceedings of the International Astronomical Union, 6, 218

\bibitem[\protect\citeauthoryear{{Spiegel} \& {Weiss}}{{Spiegel} \&
  {Weiss}}{1980}]{SpiegelWeissNature1980}
{Spiegel} E.~A.,  {Weiss} N.~O.,  1980, Nature, 287, 616

\bibitem[\protect\citeauthoryear{{Tao}, {Proctor} \& {Weiss}}{{Tao}
  et~al.}{1998}]{Tao1998}
{Tao} L.,  {Proctor} M.~R.~E.,    {Weiss} N.~O.,  1998, MNRAS, 300, 907

\bibitem[\protect\citeauthoryear{{Tobias} \& {Weiss}}{{Tobias} \&
  {Weiss}}{2007}]{TobiasWeiss2007}
{Tobias} S.,  {Weiss} N.,  2007, in {D.~W.~Hughes, R.~Rosner, \& N.~O.~Weiss}
  ed., The Solar Tachocline {The solar dynamo and the tachocline}.
p.~319

\bibitem[\protect\citeauthoryear{{Tobias}, {Brummell}, {Clune} \&
  {Toomre}}{{Tobias} et~al.}{1998}]{Tobias1998}
{Tobias} S.~M.,  {Brummell} N.~H.,  {Clune} T.~L.,    {Toomre} J.,  1998, ApJL,
  502, L177

\bibitem[\protect\citeauthoryear{{Tobias}, {Brummell}, {Clune} \&
  {Toomre}}{{Tobias} et~al.}{2001}]{Tobias2001}
{Tobias} S.~M.,  {Brummell} N.~H.,  {Clune} T.~L.,    {Toomre} J.,  2001, ApJ,
  549, 1183

\bibitem[\protect\citeauthoryear{{Tobias} \& {Hughes}}{{Tobias} \&
  {Hughes}}{2004}]{TH2004}
{Tobias} S.~M.,  {Hughes} D.~W.,  2004, ApJ, 603, 785

\bibitem[\protect\citeauthoryear{{Vasil} \& {Brummell}}{{Vasil} \&
  {Brummell}}{2008}]{VB2008}
{Vasil} G.~M.,  {Brummell} N.~H.,  2008, ApJ, 686, 709

\bibitem[\protect\citeauthoryear{{Vasil} \& {Brummell}}{{Vasil} \&
  {Brummell}}{2009}]{VB2009}
{Vasil} G.~M.,  {Brummell} N.~H.,  2009, ApJ, 690, 783

\bibitem[\protect\citeauthoryear{{Vergassola} \& {Avellaneda}}{{Vergassola} \&
  {Avellaneda}}{1997}]{VA1997}
{Vergassola} M.,  {Avellaneda} M.,  1997, Physica D Nonlinear Phenomena, 106,
  148

\bibitem[\protect\citeauthoryear{{Weiss}, {Thomas}, {Brummell} \&
  {Tobias}}{{Weiss} et~al.}{2004}]{Weiss2004}
{Weiss} N.~O.,  {Thomas} J.~H.,  {Brummell} N.~H.,    {Tobias} S.~M.,  2004,
  ApJ, 600, 1073

\bibitem[\protect\citeauthoryear{{Wissink}, {Hughes}, {Matthews} \&
  {Proctor}}{{Wissink} et~al.}{2000}]{Wissink2000}
{Wissink} J.~G.,  {Hughes} D.~W.,  {Matthews} P.~C.,    {Proctor} M.~R.~E.,
  2000, MNRAS, 318, 501

\bibitem[\protect\citeauthoryear{Zeldovich}{Zeldovich}{1957}]{Zeldovich1957}
Zeldovich Y.~B.,  1957, Sov. Phys. JETP

\end{thebibliography}
\bibliographystyle{mn2e}
\label{lastpage}
\end{document}